\documentclass[11pt]{amsart}
\usepackage{geometry}                
\usepackage{epstopdf}
 \usepackage[dvips]{epsfig}
\usepackage{graphicx}				
	
\usepackage{amssymb,amsmath}
\usepackage{gensymb} 
\usepackage[T1]{fontenc}
\usepackage[utf8]{inputenc}
\usepackage{authblk}

\usepackage{color, xcolor,bm}

\begin{document}

\begin{flushleft}
{\Large
\textbf\newline{Fluid dynamics of diving wedges}
}
\newline
\\
Lionel Vincent$^1$, Tingben Xiao$^1$, Daniel Yohann$^1$, Sunghwan Jung$^3$ and Eva Kanso$^{1,2}$\textsuperscript{*}
\\
\bigskip
1. Aerospace and Mechanical Engineering, \\
University of Southern California, Los Angeles, California, USA\\
\vspace{0.25 cm}
2. Center of Computational Biology, Flatiron Institute \\
Simons Foundation, New York, NY 10010, USA \\
\vspace{0.25 cm}
3. Department of Biomedical Engineering and Mechanics, \\
Virginia Tech, Blacksburg, VA 24060, USA\\
\vspace{0.25 cm}
* kanso@usc.edu

\end{flushleft}


\section*{abstract}
Diving induces large pressures during water entry, accompanied by the creation of cavity and water splash ejected from the free water surface. To minimize impact forces, divers streamline their shape at impact. Here, we investigate the impact forces and splash evolution of diving wedges as a function of the wedge opening angle. A gradual transition from impactful to smooth entry is observed as the wedge angle decreases. After submersion, diving wedges experience significantly smaller drag forces (two-fold smaller) than immersed wedges.  Our experimental findings compare favorably with existing force models upon the introduction of empirically-based corrections. We experimentally characterize the shapes of the cavity and splash created by the wedge and find that they are independent of the entry velocity at short times, but that the splash exhibits distinct variations in shape at later times. We propose a one-dimensional model of the splash that takes into account gravity, surface tension and aerodynamics forces. The model shows, in conjunction with experimental data, that the splash shape is dominated by the interplay between a destabilizing Venturi-suction force due to air rushing between the splash and the water surface and a stabilizing force due to surface tension. Taken together, these findings could direct future research aimed at understanding and combining the mechanisms underlying all stages of water entry in application to engineering and bio-related problems, including naval engineering, disease spreading or platform diving.

\section{Introduction}\label{sec:introduction}


We investigate the motion of a rigid wedge diving across an air-water interface. 
Water entry problems  have appealed to scientists and engineers alike for more than a century. 
The beauty of splashes were first examined using high-speed photography by \cite{Worthington1908}, and were later studied in the context of naval engineering problems  \cite{VonKarman1929,Wagner1932}. While naval-oriented research is still very active \cite{Abrate2011}, understanding and predicting forces on entering objects is also relevant for other fields, such as air/water missiles \cite{May1952}, aerospace engineering \cite{Seddon2006}, diving birds \cite{Jung2016,RopertCoudert2004}, lizard locomotion \cite{HsiehLauder2004}, prevention of injury in olympic diving \cite{Harrison2012}, and dissemination of seeds \cite{Amador2013}, aroma \cite{Ghabache2014}, and diseases \cite{Bourouiba2015,Joung2017}.

\begin{figure}
\centering
\includegraphics[width=0.99\textwidth]{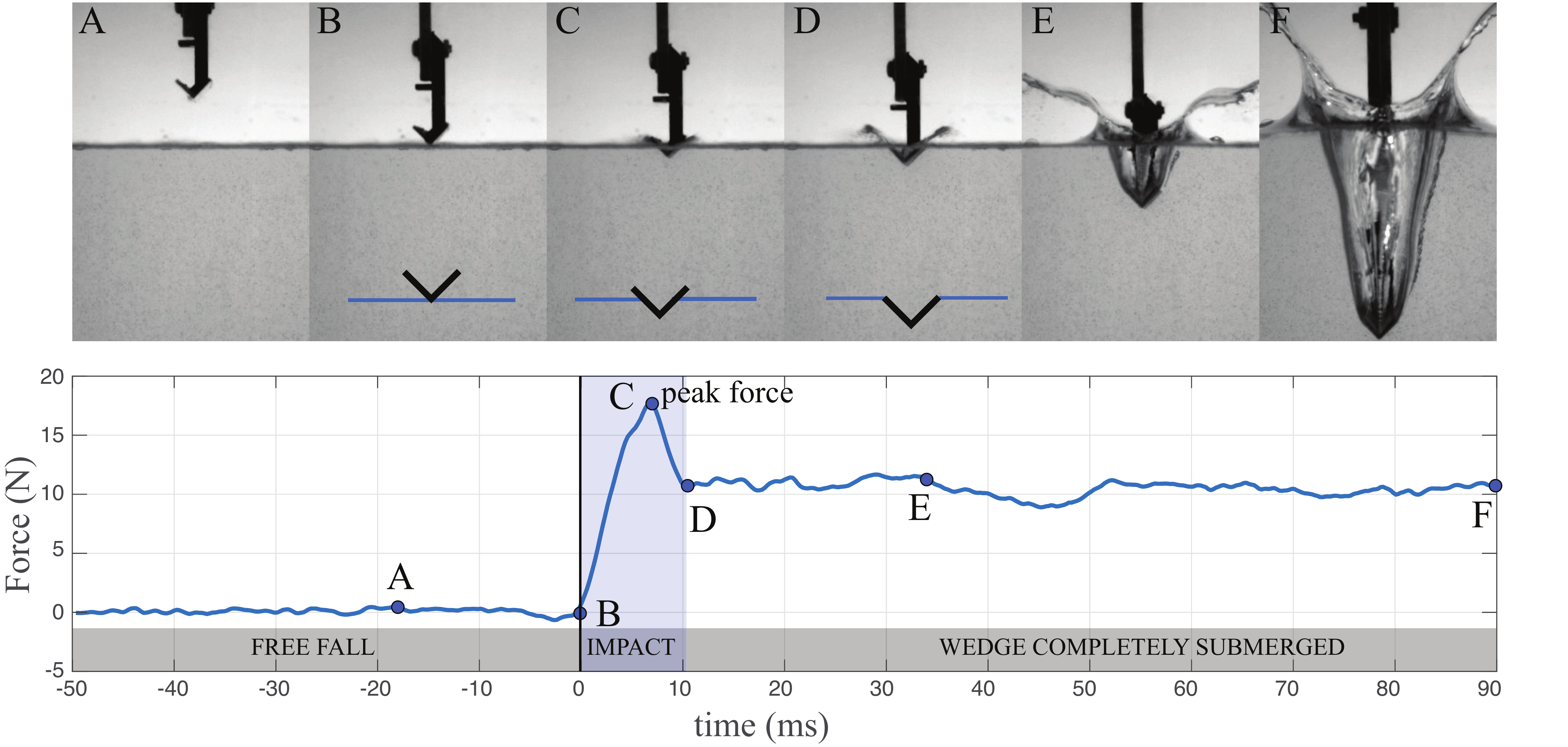}
\caption{Typical water entry sequence and force measurement, for a $\alpha = 90 \degree$ wedge impacting at $V= 1.70$ m/s. For large enough wedge angles ($\alpha > 70 \degree$), the compression force peaks before the wedge is fully submerged; a smaller, quasi-constant force is observed after submersion. $t = 0$ is defined as the first contact between the wedge and the water surface. The entry generate a thin and fast-moving splash, likely to interact with the surrounding air. }
\label{fig_gendescription}
\end{figure}

In this study, wedges of width $d$ and opening angle $\alpha$ are dropped under gravity $g$ from a height $H$. They reach the air/water interface with a velocity $V \approx \sqrt{2 g H}$. Figure~\ref{fig_gendescription} shows a typical sequence of events following water entry, and the corresponding vertical force acting on the wedge's supporting arm. Here, the entry velocity is $V = 1.70$ m/s, and the wedge's width is $d = 18$ mm. This sequence illustrates several generic features of water entry. Shortly after first contact with the water surface, the vertical force quickly rises from 0 to a peak value (B-C). 
The peak is very prominent and is reached before the wedge is completely submerged. This characteristic pattern is called ``slamming''. The vertical force then decreases  (C-D) and changes little once the wedge is fully submerged (D-E). Following the wedge's submersion (D-E-F), two visually-striking events occurs. First, an air-filled expanding cavity is created in the wedge's wake. Second, a curved splash is ejected upwards and sideways from the point of impact. The ejection velocity can be significantly higher than the wedge's entry velocity $V$: the thin and fast-moving splash is subject to aerodynamic interactions leading to non-trivial arabesques (E). This work is focused on the description and modeling of the vertical force generated during entry, as well as modeling the splash sheet kinematics. The first part deals with early stages of water entry (prior to wedge's total submersion), and in particular the transition from smooth to impactful entry as the wedge angle increases. The second part analyzes, using a combination of empirical observations and low-order physics-based models,  the long term evolution of the splash projected upward from the edges of the wedge during and following water entry.
 
\subsection{Impact forces on objects entering water}

The fluid dynamics literature is divided into two types of studies: those that focus on slamming forces and other that focus on the cavity and splash  formation. \cite{VonKarman1929} was the first to estimate the slamming forces on wedges based on conservation of fluid-wedge momentum and the added mass effect. Shortly after, \cite{Wagner1932} presented a refined model, also based on potential flow theory, in which free surface elevation was taken into account. \cite{Zhao1993} and \cite{Zhao1996} were the first to propose a fully nonlinear solution for the coupled fluid-wedge system, and \cite{Mei1999} developed a conformal mapping technique that built upon these findings. In a series of publications \cite{Korobkin1988,Korobkin1996,Molin2001,Korobkin2004}, Korobkin offered a number of insightful analytical models of fluid entry under various conditions, including impact of a rigid body with an attached cavity and of a perforated wedge. Recently, explicit finite element methods have been employed by \cite{Bereznitski2001}, \cite{Stenius2006}, \cite{Wang2012} and \cite{Sotiropoulos2014} to predict slamming loads. While many theoretical and numerical models predict impact forces on wedges of relatively small angles, there are few experimental studies that seek to validate these predictions~\cite{Wu2004,Yettou2006,Zhao1996,Tveitnes2008}. Even fewer studies consider and quantify the unavoidable three-dimensional effects of real-life experiments \cite{Zhao1996}. 

In the first part of this paper, we characterize the forces acting on the wedge during water entry (before and after submersion), and we compare our experimental results to existing theories. We show that wedges of large angle undergo ``impactful'' entry, because of a large transient peak force felt before submersion. In contrast, low-angle wedges enter smoothly, with a force gradually rising from zero to a terminal value. The transition between impactful and smooth entry can be predicted by a clever use of existing data and theory. We also show that the drag force acting on the wedge after submersion is quasi-constant, and because of the presence of the cavity, it is significantly lower than the drag force of an immersed wedge . 

\subsection{Splash and cavity evolution}

The study of splashes and cavities date back to the beginning of high-speed photography \cite{Worthington1908} and continues to be the topic of numerous publications \cite{BirkhoffZarantonello1957,DuclauxClanet2007,AristoffBush2009}. Most studies consider either the dynamics of the cavity or the development of the splash.  Cavities evolve relatively slowly with resptect to the objet's velocity and are thus easier to visualize and analyze. The retroaction of the cavity dynamics on the trajectory of the impacting object is of interest for the military, in order to make bullets or air/water anti-torpedo missiles reach an underwater target \cite{BirkhoffZarantonello1957,May1952}. At very large impact velocity, low-pressure area in the object's wake triggers cavitation \cite{Truscott2014}, which affects the stability of the object's trajectory. At slower impact velocity, the cavity is primarily a result of inertia. Many research contributions focus on describing, classifying, and modeling the cavity created by various impactors \cite{DuclauxClanet2007,AristoffBush2009,Truscott2009,MarstonThoroddsen2016}. 
Splahes, in contrast to cavities, are thin fast developing features of elusive nature, and, unlike cavities, they are strongly dependent on the geometry of the object. Splashes have been extensively studied in the canonical problem of impacting spheres. Water entry of spheres is usually accompanied by a nearly vertical splash curtain \cite{Worthington1908,AristoffBush2009}. Entry of wedges, because of the large horizontal momentum imparted to the liquid, generally induce splashes curving outwards and downwards \cite{Greenhow1987}. These splashes are generally better defined, thicker, and less sensitive to wettability than splashes created by round objects, but they are considerably less studied than the latter.

For splashes, perhaps more than for any other aspect of water entry, the devil is in the details. As \cite{DuezClanet2007} pointed out, the traditional vision of a water entry driven by inertia for high-enough entry velocity is essentially wrong. They demonstrated that capillary effects such as wettability have first-order effect on the cavity formation and subsequent splashes for spheres. Likewise, \cite{MarstonThoroddsen2016} showed that capillary wrinkles originating from the contact line on the sphere are responsible for the dramatic splash shape upon ``buckling''. Viscosity is also traditionally thought to be irrelevant.
Lastly, the surrounding air  affects various aspects of the splash, including the ejecta \cite{Quere2005,Xu2005}, cavity collapse in axisymetric problems \cite{GekleGordilloLohse2010,Gordillo2005} and quasi-two-dimensional problems \cite{Wang2015}, and surface seal \cite{MarstonTruscott2015}.


Attempts to simulate splashes have been met so far with mixed results, including for simple geometries such as wedges. In early models and simulations, the jet is either not considered at all    \cite{Wagner1932,Logvinovich1972}, or cut-off when leaving the wedge to avoid numerical difficulties  \cite{Zhao1996,Battistin2003}.  Recent contributions use smoothed particle hydrodynamics \cite{Kai2009}, level-set immersed boundary methods \cite{Sotiropoulos2014} or boundary element methods based on potential flow theory \cite{Wu2004,Bao2016} to solve the full system and account for the free jet. However, most simulations either misrepresent  the splash development \cite{Sotiropoulos2014} or fail to include potentially important parameters such as surface tension \cite{Bao2016}. In short, existing descriptions of the splash shape lack a proper framework to help understand the effects of various physical forces on the splash evolution.



The second part of this study focuses on the cavity and splash evolution created by diving wedges. Using high-speed photography to reconstruct the shapes of the cavity and splash, we show that the cavity is self-similar for various entry velocities but the splash is not.
In order to investigate the physics underlying the splash evolution, we develop a one-dimensional model based on the idea that the splash is primarily ballistic, and can be represented by a succession of discrete particles ejected from the free water surface.  The model shows, in conjunction with the experimental data, that the splash shape is the result of the interplay between aerodynamic interactions that favor bending and capillary effects that tend to cancel curvature. 

\begin{figure}
\centering
\includegraphics[width=0.99\textwidth]{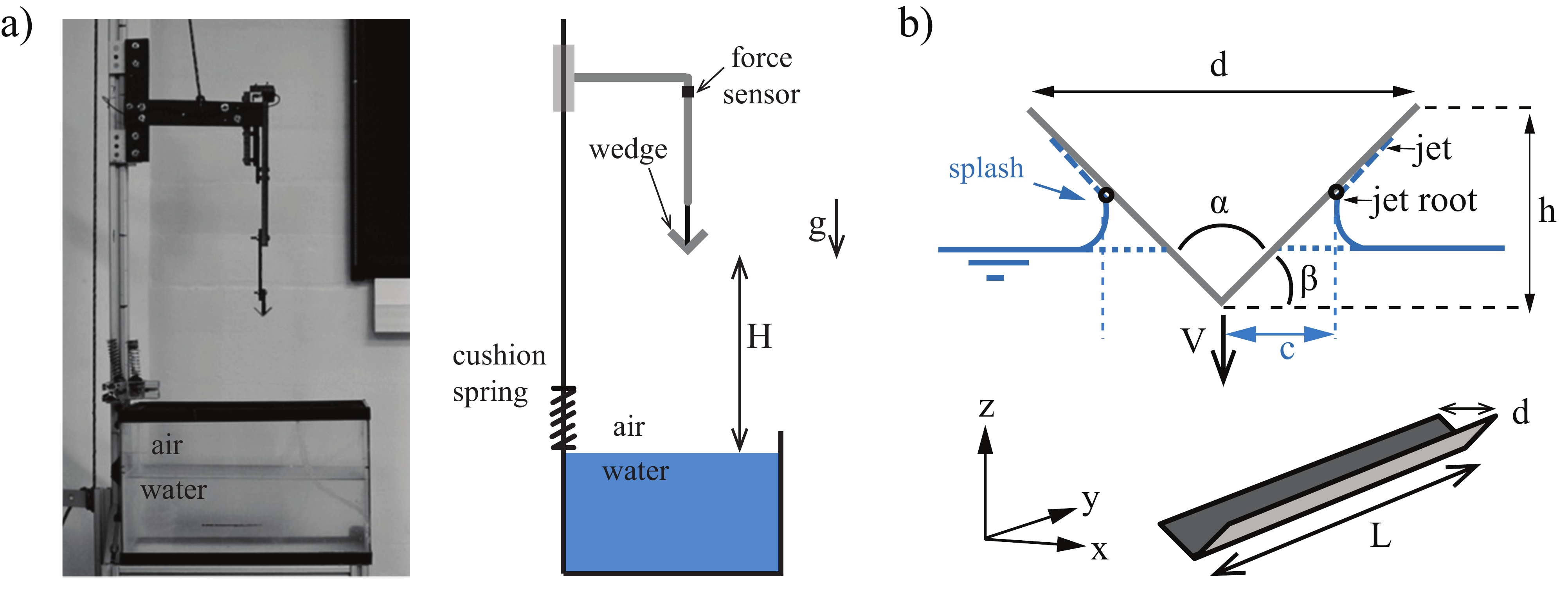}
\caption{Experimental setup.}
\label{fig_setup}
\end{figure}

\section{Experimental setup}

All experiments were driven by gravity and were performed in a water-filled rectangular acrylic tank measuring 51$\times$26$\times$32 cm$^3$. To ensure straight and reproducible entry conditions, a vertical tower 1.5 m in height was used to guide the falling wedge, see figure~\ref{fig_setup}(a). Wedges dropped from a height $H$ reach the air/water interface with a velocity $V \approx \sqrt{2 g H}$. By varying $H$, the entry velocity $V$ could be chosen in the range from 0 to 3.5 m/s. 
The wedges were 3D printed using Hatchbox PLA plastic of dimensional accurary $\pm 0.05$ mm. Two sets of wedges were printed: in the first set,  the opening angle $\alpha$ of the wedge was varied from $60 \degree$ to $120 \degree$ while the wedge width $d = 36$ mm and length $L = 150 $ mm were held constant, yielding an aspect ratio $L/d = 4.2$. In the second set,  the aspect ratio was fixed at  $L/d = 4.6$ while the projected area $L\times d$ was varied in the range $1400 - 5940$ mm$^2$. Additionally, $90 \degree$ wedges of a smaller width $d=18$ mm but same aspect ratio $L/d$ were used to check that several quantities were independent of $d$. The mass of all moving parts $M$ of the wedge and drop mechanism was around 875 g, with the mass of the wedge itself ranging between 15 and 30 grams. The weight $M g$ was chosen so that it balances the average drag and impact forces during water entry, thus limiting acceleration or deceleration during water entry and allowing us to treat the velocity $V$ of the wedge as constant throughout water entry.

\begin{figure}
\centering
\includegraphics[width=0.92\textwidth]{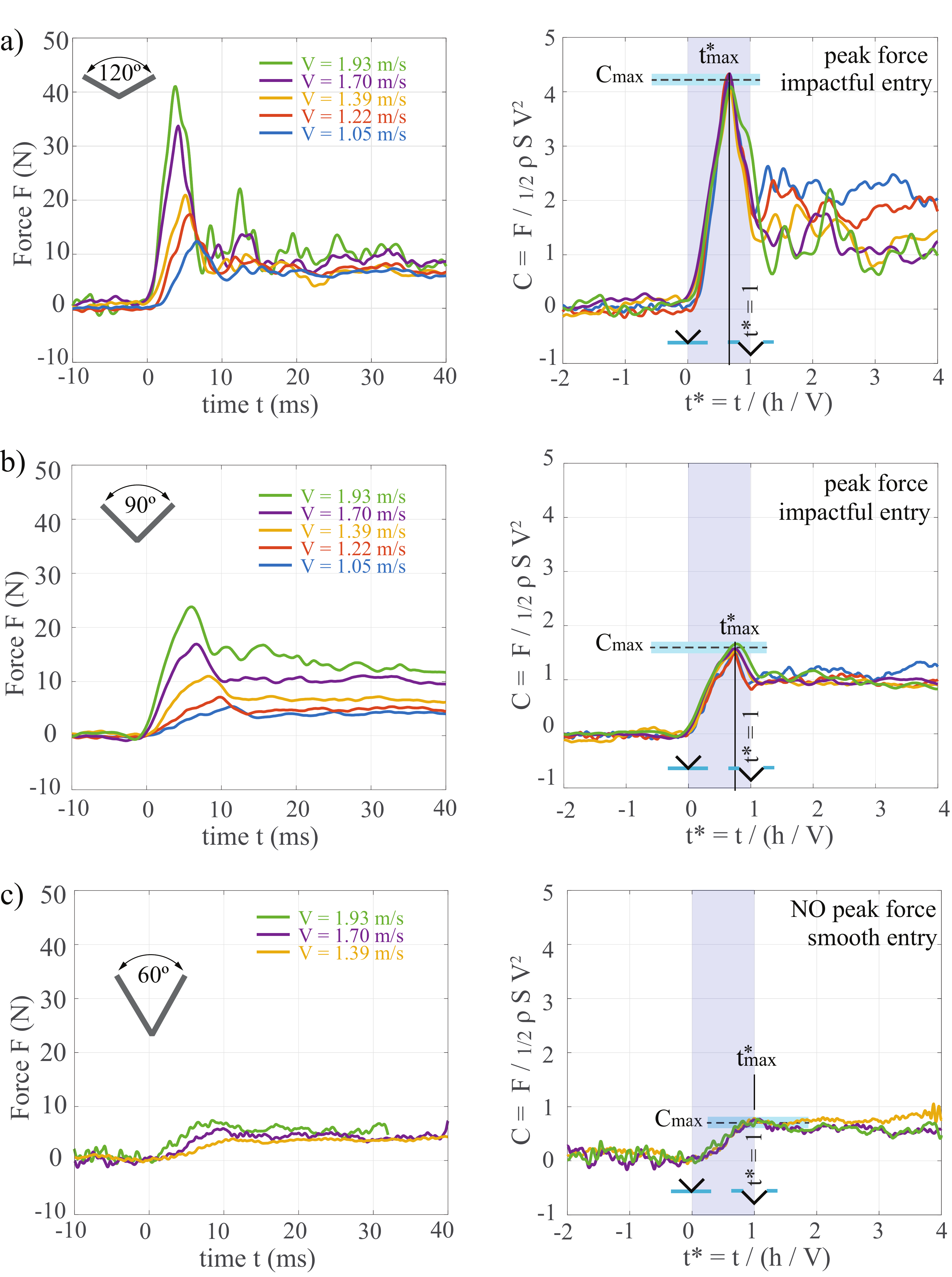}
\caption{Dimensional (left) and non-dimensional (right) force history during water entry for $120 \degree$, $90 \degree$, and $60 \degree$ wedges at various speeds. Note the almost perfect collapse of the force coefficient during the early stages of entry (right). A prominent peak force appearing for $t^* < 1$ is observed for $\alpha = 120 \degree$ and $\alpha = 90 \degree$, which is considered impactful entry; for smaller angles such as (c), the measured force gradually rises from $t^*=0$ (first contact) to $t^*=1$ (wedge fully submerged): the entry is smooth. $L = 150$ mm, $d = 36$ mm.}
\label{Fig:Force_and_Cp}
\end{figure}

The water impact problem can be described by three non-dimensional parameters: the Weber number $\textrm{We} = \rho d V^2 / \sigma$, where $\rho = 1000$ Kg/m$^3$ is the water density and $\sigma = 70 \cdot 10^{-3}$ N/m is the surface tension of the air/water interface, the Froude number $\textrm{Fr} = V / \sqrt{g d}$, where $g = 9.81$ m/s$^2$ is the gravitational constant, and the Reynolds number $\textrm{Re} = d V / \nu$, where $\nu = 10^{-6}$ m$^2$/s is the kinematic viscosity of water. 
Given the experimental parameters, Weber numbers $\textrm{We}$ were in the range 200$-$2000, indicating fluid inertia should dominate surface tension effects. The  Froude number $\textrm{Fr}$ ranged from 1.7 to 5 in the present experiments, indicating that hydrostatic effects may have a noticeable contribution after the wedge is fully submerged. 
Lastly, the Reynolds number $\textrm{Re}$ ranged between $10^4$ and $10^5$, indicating that effects of viscosity can safely be ignored. It is worth noting that the relative influence of gravity with respect to surface tension can be indicated by the Bond number $\textrm{Bo} =$ We/Fr$^2$ $= \rho g d^2 / \sigma$,  which ranged from 45 to 180 in our experiments, confirming surface tension should not be a major player in the forces acting on the wedge during water entry (but will be important in splash development). 

Forces acting on the wedge were measured directly by mean of a compression load cell (FC22, Measurement Specialties), secured between the horizontal arm connected to the sliders and the vertical rigid stem on which the wedge is mounted (see figure~\ref{fig_setup}(b)). An NI PCIe 6323 acquisition card was used to read data from the compression sensor at a rate of 8000 Hz. A short spring was used to keep the wedge and stem in contact with the sensor during the free-fall. This small elastic force was subtracted from the presented measurements. 


High speed photography was used to characterize jet and air cavity development. Water entries were recorded at frame rates ranging between 1600 and 9000 fps using a Phantom Miro M-110 high speed camera. A minimum of five videos from ten experiments were obtained for each wedge at each dropping height. Films captured drops from the front view, side view, and perspective view.


\section{Forces during water entry}

We characterize the forces acting on diving wedges, before and after submersion, with particular emphasis on the gradual transition from impactful to smooth water entry as the wedge angle decreases. 

\subsection{Force measurements}


Figure~\ref{Fig:Force_and_Cp}, left panel, shows the force evolution recorded by the force sensor for three wedges $\alpha = 120 \degree$, $\alpha = 90 \degree$ and $\alpha = 60 \degree$, at various entry velocities $V$. As in figure~\ref{fig_gendescription}, the force quickly rises from 0 to a positive value after impact.  For $\alpha = 120 \degree$ and $\alpha = 90 \degree$, the force peaks after entry, with peak values in the range of $10 - 40$ N, then decreases to a quasi-constant value once the wedge is fully submerged. For $\alpha = 60 \degree$, the force gradually increases from 0 to a terminal value: no prominent peak force is observed. We call the first pattern ``impactful entry'' and the second ``smooth entry''.

Given the inertial flow regime ($\textrm{Re} > 10^4$), we scale force by $\frac{1}{2} \rho S V^2$, where $S = L \times d$ is the projected area of the wedge and let $C = F / \frac{1}{2} \rho S V^2$ denote the non-dimensional force coefficient. We scale time by the inertial time scale $h / V$ such that $t^* = t / (h/V)$ is non-dimensional time, with $t^* = 0$ being the time of impact and $t^* = 1$ corresponding to the instant when the wedge is fully submerged with respect to the undisturbed water surface. 
Dimensionless data is depicted in the right panel of figure~\ref{Fig:Force_and_Cp}. At all wedge angles, the data for different entry velocities collapses very well for $t^* \leq 1$, confirming that the force and time scales are appropriate. From these graphs, we extract a single value of the maximum force $C_{\textrm{max}}$. We also mark the time $t^*_{\textrm{max}}$ when the maximum force happens, around $t^*_{\textrm{max}}=0.66, \, 0.73, \, 1$ for $\alpha = 120 \degree, \, 90 \degree, \, 60 \degree$, respectively.
For $t^* > 1$, the force measurements do not collapse. We observe larger force coefficients at smaller entry velocities, particularly obvious for $\alpha = 120 \degree$, which can be attributed to hydrostatic forces becoming significant as $\textrm{Fr}$ approaches unity.

\begin{figure}
\centering
\includegraphics[width=1.0\textwidth]{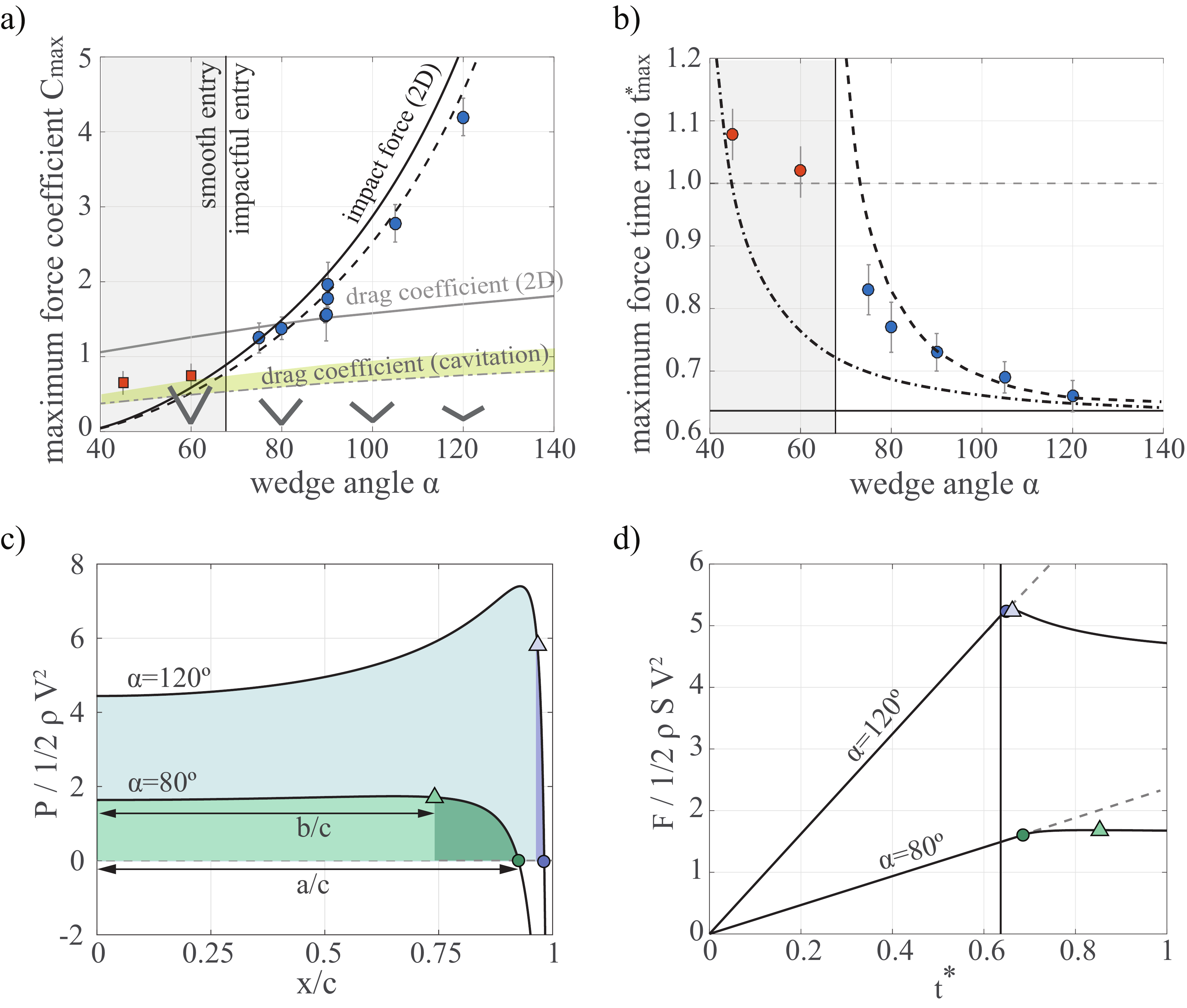}
\caption{(a) The maximum force coefficient increases sharply with wedge angle $\alpha$, and follows theoretical predictions by \cite{Logvinovich1972} (solid line); excellent agreement is observed when Logvinovich's two-dimensional theory is corrected for  finite aspect ratio $L/d=4.2$ (dashed line). (b) For $\alpha \leq 70 \degree$, the maximum force happens roughly when the wedge is fully submerged ($t_{\textrm{max}}^* \approx1$), while for  $\alpha > 70 \degree$, $t_{\textrm{max}}^*$ drops sharply and converges to the naive prediction from Wagner's theory $t^*_{max} = 2/\pi$. (c-d) The theoretical pressure profile can be used to refine the prediction for $t^*_{max}$. Assuming the maximum force occurs when the negative pressure area escapes the wedge ($a=d/2$, circles) lead to an analytical, refined estimation (dash-dotted line in (b)). In reality, the pressure integral (force $F$) on the wedge no longer increases when $b = d/2$ (triangles), leading to a second numerical prediction (dashed line in (b)), providing the same trend as the experimental measurements. Transition from smooth to impactful entry is expected when the expected 2D drag coefficient equals the theoretical impact force ($\alpha \approx 79 \degree$). 2D drag data from \cite{Hoerner1965}. In (a) and (b), red square data points denote smooth entry, defined as the absence of prominent peak. }
\label{Fig:Cp_vs_alpha}
\end{figure}

%

These measurements are repeated with wedge angles $\alpha=$ 60, 75, 80, 90, 105 and 120$\degree$, leading to the values of $C_{\textrm{max}}$ and $t^*_{\textrm{max}}$ shown in figure~\ref{Fig:Cp_vs_alpha}(a) and (b) and table~\ref{tab:Cp}. Rather intuitively, $C_{\textrm{max}}$ increases sharply with $\alpha$, starting at $C_{\textrm{max}} = 0.65$ for $\alpha = 45 \degree$ and reaching $C_{\textrm{max}} = 4.2$ for $\alpha = 120 \degree$, while $t^*_{\textrm{max}}$ decreases with $\alpha$, starting at $t^*_{\textrm{max}} = 1.08$ for $\alpha = 45 \degree$ and plummetting to $t_{\textrm{max}}^* = 0.66$ for $\alpha = 120 \degree$. No prominent peak force was observed for wedges with $\alpha < 75 \degree$. 



\subsection{Force model}
Most models of the impact force during water entry are based on Wagner's original work \cite{Wagner1932}, which provides good predictions of the impact force for nearly flat wedges ($\alpha \approx180 \degree$), but lead to large overestimates at moderate opening angles (close to a 2-fold error for $\alpha = 120 \degree$). Among the models that improved upon Wagner's theory, Logvinovich's model \cite{Logvinovich1972} seems to provide the most accurate description of the maximum force value, while also providing a good prediction of the pressure on the wedge \cite{Korobkin2004,Panciroli2015}. Here, we show that (i) our experimental data compares favorably with Logvinovich's model once 3D effects are taking into account and (ii) the transition between smooth and impactful entry, occuring for $\alpha \approx 70 \degree$, can be anticipated using the same model.


Logvinovich's model is based on potential flow theory and uses asymptotic expansions around large wedge angles ($\alpha \approx 180 \degree$)  to solve for the flow velocity at the wedge, leading to an approximate pressure distribution along the wedge as a function of the horizontal coordinate $x$ and time $t$,
\begin{equation}
P(x,t) = \frac{1}{2} \rho V^2 \left[ \frac{\pi}{\tan{\beta}} \frac{c}{\sqrt{c^2-x^2}} - \frac{c^2}{c^2-x^2} \right].
 \label{eq:P_Log}
\end{equation}
Here, the deadrise angle  $\beta = (\pi - \alpha)/2 $ is small, and $c$ designates the jet root corresponding to the full extent of the wetted region in this model (see figure~\ref{fig_setup}). The evolution of $c$ is given by the so-called ``Wagner condition''
\begin{equation}
c(t) = \pi V t / (2 \tan{ \beta}), \label{eq:c}
\end{equation}
where $V t$ is the wedge's penetration distance from the undisturbed water surface. Our experimental measurements of $c(t)$  (see section~\ref{sec:splash_ejection}) agree well with this prediction. The non-dimensional pressure $P(x,t) / \frac{1}{2}\rho V^2$ based on~\eqref{eq:P_Log} is depicted  in figure~\ref{Fig:Cp_vs_alpha}(c) as a function of $x/c$.  It is positive on most of the wetted region of the wedge but becomes negative close to the periphery, and tends to $-\infty$ for $x \to c$. The sign change happens at $a/c = \sqrt{1-(\tan{\beta} / \pi)^2}$ (marked by circular symbols in figure \ref{Fig:Cp_vs_alpha}(c)). This singularity is common to all Wagner-based models and is usually regularized using the anzatz that only positive pressure matters \cite{Korobkin2004}: pressure should be integrated from $x=0$ to $x=a$, which yields (accounting for both sides of the wedge)
\begin{equation}
F(t) = 2 \int_0^{a(t)} P(x,t) dx = \rho V^2 c(t) \left[ \frac{\pi^2}{2 \tan{\beta}} - K(\beta) \right],  \label{eq:F}
\end{equation}
where
\begin{equation}
K(\beta) = \frac{\pi}{\tan{\beta}} \left[\frac{\pi}{2} - \arcsin{(a/c)} \right] - \frac{1}{2} \ln{\left[ \frac{1+a/c}{1-a/c} \right]} . \label{eq:K_for_F}
\end{equation}
To predict the maximum force, we assume that it is reached when the wetted length $c$ equals the wedge's half-width~$d/2$, that is to say, when the splash root escapes the wedge. From (\ref{eq:c}), one gets $t^*_{\textrm{max}} = 2 / \pi \approx 0.636$, independently of $\alpha$.
Substituting the prediction for $t^*_{\textrm{max}}$ into (\ref{eq:F},\ref{eq:K_for_F}) leads to 
\begin{equation}
F_{\textrm{max}} = \frac{1}{2} \rho S V^2 \left[ \frac{\pi^2}{2 \tan{\beta}} - K(\beta) \right] .  \label{eq:F_max}
\end{equation}
In figure~\ref{Fig:Cp_vs_alpha}(a) and (b), we compare $F_{\textrm{max}} / \frac{1}{2} \rho S V^2$  and $t^\ast_{\textrm{max}}$ (solid lines) with our experimental data.  The prediction for $t^\ast_{\textrm{max}}$  matches roughly with the experimental data for large $\alpha$, but the discrepancy increases sharply as $\alpha$ decreases. 
The force prediction is consistently (about 15\%) larger than our data for $\alpha > 90 \degree$.
One reason for this discrepancy stems from the fact that the prediction is based on two-dimensional theory 
while our wedges are not infinitely long ($L/d = 4.2$): it is thus necessary to consider three-dimensional effects. Following the suggestion of \cite{Zhao1996}, we correct Logvinovich's two-dimensional prediction using Meyeroff's results \cite{Meyerhoff1970}, who calculated the added mass coefficients of rectangular plates of various length-to-width ratio. His calculations show that in order to accurately represent the effect of finite aspect ratio in situations where added mass plays an important role, two-dimensional predictions have to be corrected by a factor approximately equal to $1 -  d/(2L)$. In our case, this would predict a 12\% decrease in maximum force (dashed line in figure~\ref{Fig:Cp_vs_alpha}(a)), which is in good agreement with the experimental data for moderate and large wedge angles.

\begin{table}
\centering  %
\begin{tabular}{c|cccc}  %
$\alpha$  & $C_{\textrm{max}}$   & $C_{\textrm{max}}$(theory)  & $t^*_{\textrm{max}}$ & $t^*_{\textrm{max}}$ (theory)  \\ \hline
$45 \degree$    & $0.65\pm0.3$ & $0.15$   & $1.08\pm0.04$   & $-$   \\
$60 \degree$    & $0.75\pm0.3$ & $0.58$   & $1.02\pm0.04$   & $-$   \\
$75 \degree$    & $1.25\pm0.3$ & $1.24$   & $0.83\pm0.07$   & $0.947$   \\
$80 \degree$    & $1.38\pm0.3$ & $1.75$  & $0.77\pm0.09$   &  $0.833$ \\
$90 \degree$    & $1.7 \pm0.3 $ & $2.08 $ & $0.73\pm0.3$     & $0.734$   \\
$102 \degree$  & $2.78\pm0.3$ & $3.25$  & $0.69\pm0.025$   & $0.686$   \\
$120 \degree$  & $4.2\pm0.3$   & $5.15$  & $0.66\pm0.025$ & $0.656$ 
\end{tabular}
\caption{Measured and theoretical force coefficient $C_{\textrm{max}}$ and time ratio $t^*_{\textrm{max}}$ (occurrence of maximum force).}
\label{tab:Cp}
\end{table}

These predictions for $F_{\textrm{max}}$ and $t_{\textrm{max}}$ are based on the assumption that the maximum force happens when $c$ equals the wedge's half-width~$d/2$. 
Yet, we know that the theoretical pressure profile along the wetted length (figure~\ref{Fig:Cp_vs_alpha}(c)) is negative close to the jet root, namely from $x = a$ to $x=c$. 
To refine the prediction of $t_{\textrm{max}}$, we consider the maximum force to happen when the region of negative pressure has escaped the wedge, that is, when $a = d/2$, which yields $t^*_{\textrm{max}} = (2/\pi)(c/a) = (2/\pi) / \sqrt{1-(2 \tan{\beta} / \pi)^2}$.
This improved prediction, represented in figure~\ref{Fig:Cp_vs_alpha}(b) as a dash-dotted line,  increases sharply as $\alpha$ decreases but  underestimates the experimental results, especially for moderate wedge opening angles. To refine the prediction further, we note that the integral of the pressure profile on the wedge continues to increase until $b= d/2$, as shown in figure~\ref{Fig:Cp_vs_alpha}(c) (triangles). Numerical computation of this integral leads to a refined estimate of $t^*_{\textrm{max}}$, represented as a dashed line in figure~\ref{Fig:Cp_vs_alpha}(b). This estimate is now in reasonable agreement with the experimental values for wedge angles down to $\alpha \approx 80 \degree$.

As the opening angle $\alpha$ of the wedge decreases, the peak force decreases. The predictions and experimental data gradually diverge for smaller $\alpha$.  For $\alpha$ below 80$\degree$, the measured maximum force is comparable to the drag coefficient on an immersed wedge.
We anticipate that transition from impactful to smooth entry as $\alpha$ decreases happens when the drag force equals the predicted impact force. Using 2D drag force data from the literature \cite{Hoerner1965}, we can estimate the transition to occur  at $\alpha = 79\degree$. This prediction is in good \emph{qualitative} agreement but overestimates the experimentally observed transition, which occurs between 60$\degree$ and 75$\degree$. One reason for this discrepancy is due to the fact that the drag force acting on diving wedges is smaller than we would expect for an immersed wedge.



\subsection{Drag force with a cavity}

\begin{figure}
   \centering
   \includegraphics[width=1.0\textwidth]{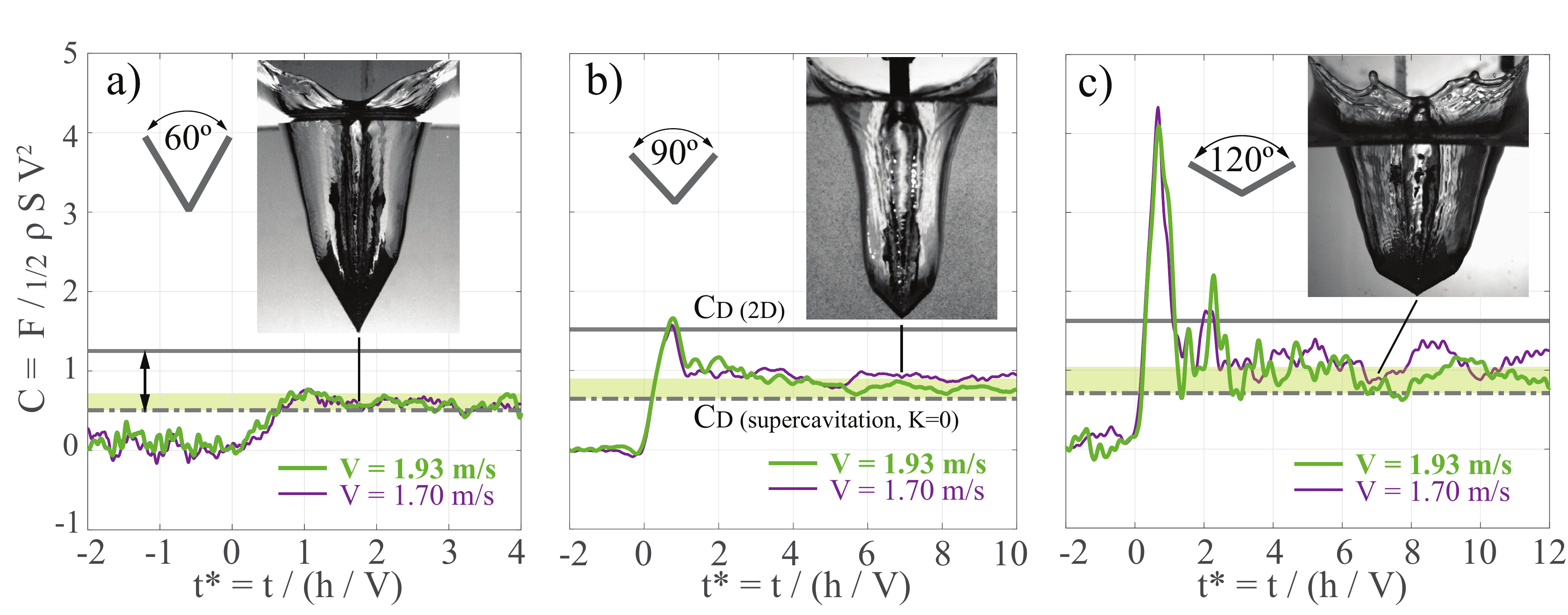} 
   \caption{Force history on wedges with opening angles $\alpha = 60, 90, 120\degree$, in the limit of small hydrostatic effects ($\textrm{Fr} > 2.8$). The non-dimensional drag force on the entering wedge is significantly smaller than the corresponding drag coefficients of an immersed wedge (solid grey line); it matches drag force measured on cavitating wedges (dash-dotted line and yellow overlay).}
    \label{fig:drag_force}
  \end{figure}

\begin{figure}
   \centering
   \includegraphics[width=1.0\textwidth]{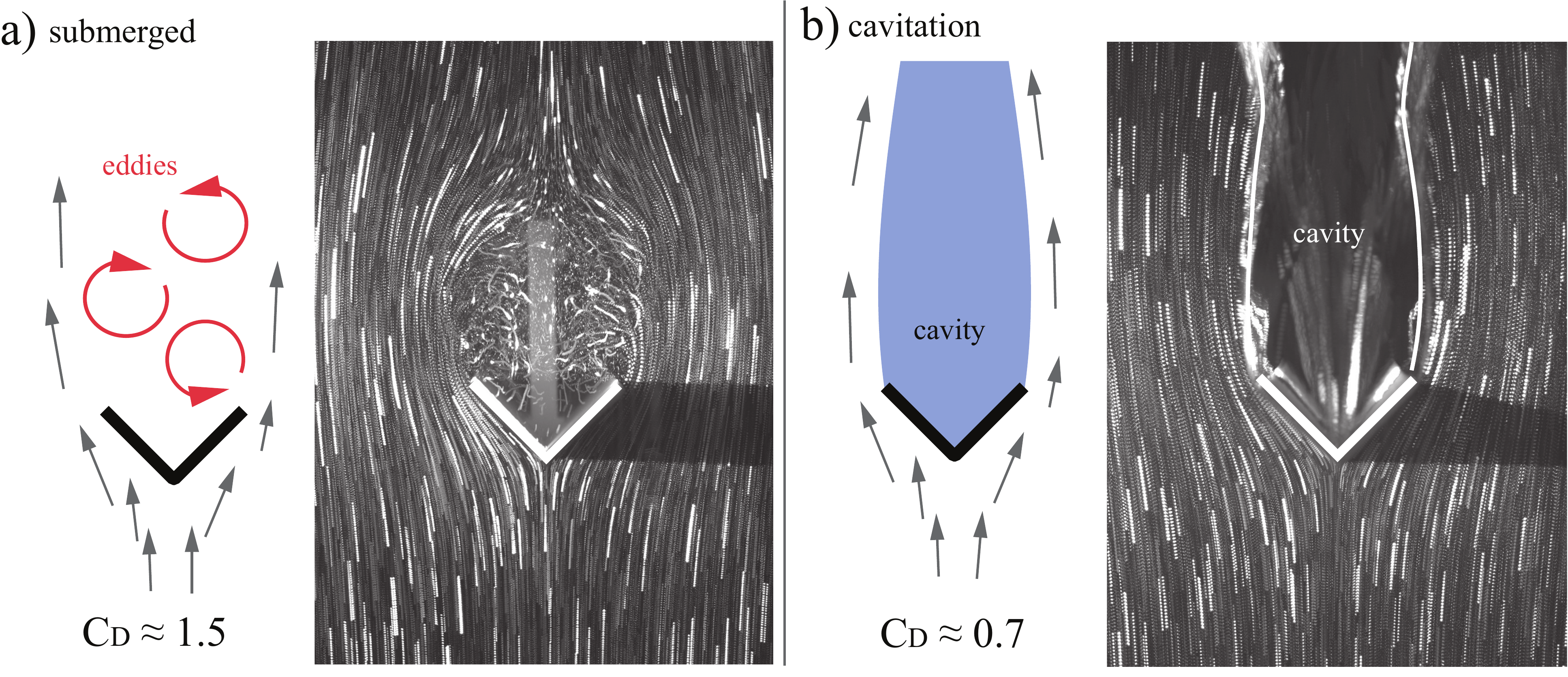} 
   \caption{Cavity and drag reduction. The presence of a transient cavity induces striking changes in the wake's wake and dramatically reduces the drag on the wedge. $\alpha = 90 \degree$, $V \simeq 1$  m/s. Equivalent exposure for both picture is $\Delta t = 12$ ms.}
    \label{fig:cavitation_and_drag}
  \end{figure}

Once the wedge is fully submerged, it reaches a quasi-permanent regime characterized by a relatively small vertical force (figure~\ref{fig:drag_force}). The measured force is significantly lower than the drag force on a fully submerged wedge of the same angle (dash-dotted line in figure~\ref{fig:drag_force}). The obvious difference between the two cases is the presence of the cavity in the wake of the diving wedge. There is no systematic model of how transient cavities affect drag on translating bodies. However, the influence of cavities on drag has been studied extensively in the context of cavitation \cite{Knapp1970}. Following a suggestion by \cite{Zhao1996}, we  compare our experimental data to cavitation results.

Results for cavitating cases show that the presence of the cavity induces a dramatic reduction of drag, between two and three folds. The reduction depends on the cavitation number $K = (p_\infty - p_{\textrm{vap}}) / \frac{1}{2} \rho V^2$, which compares the Bernouilli pressure drop $\frac{1}{2} \rho V^2$, with the pressure drop needed to reach vapor pressure $p_\infty - p_{\textrm{vap}}$, and indicating the likeliness of cavitation happening. $V$ denotes the object translational velocity, and $p_\infty$ is the pressure far away from the object. Typically, complete cavitation is reported for values up to $K \approx 0.4$, and partial cavitation happens beyond this value \cite{Knapp1970}. Because the cavities in the present work are fully formed, a sensible range of value for $K$  is between 0 and 0.4. Note that we did not arbitrarily fix $K$ to 0 as previous authors did \cite{Zhao1996}. \cite{Knapp1970} gives the following values for steady drag coefficients for supercavitating wedges at $C_D = C_{Do} (1 + K)$, with $C_{Do}$ ranging from $0.489$ to $0.745$ for $\alpha = 60 \degree$ and $\alpha = 120 \degree$. The corresponding ranges of $C_D$ are reported as a yellow zone in figure~\ref{fig:drag_force} for each wedge angle. The experimental force measurements based on the force sensors exhibit residual oscillations, which we attribute to the drop mechanism. Despite these osillations, the predictions proved to be in good agreement with our experimental data. This suggest that transient (inertial) cavities affect the drag of diving wedges the same way as cavitation does. 

\section{Splash Model}

\begin{figure}
   \centering
   \includegraphics[width=0.6\textwidth]{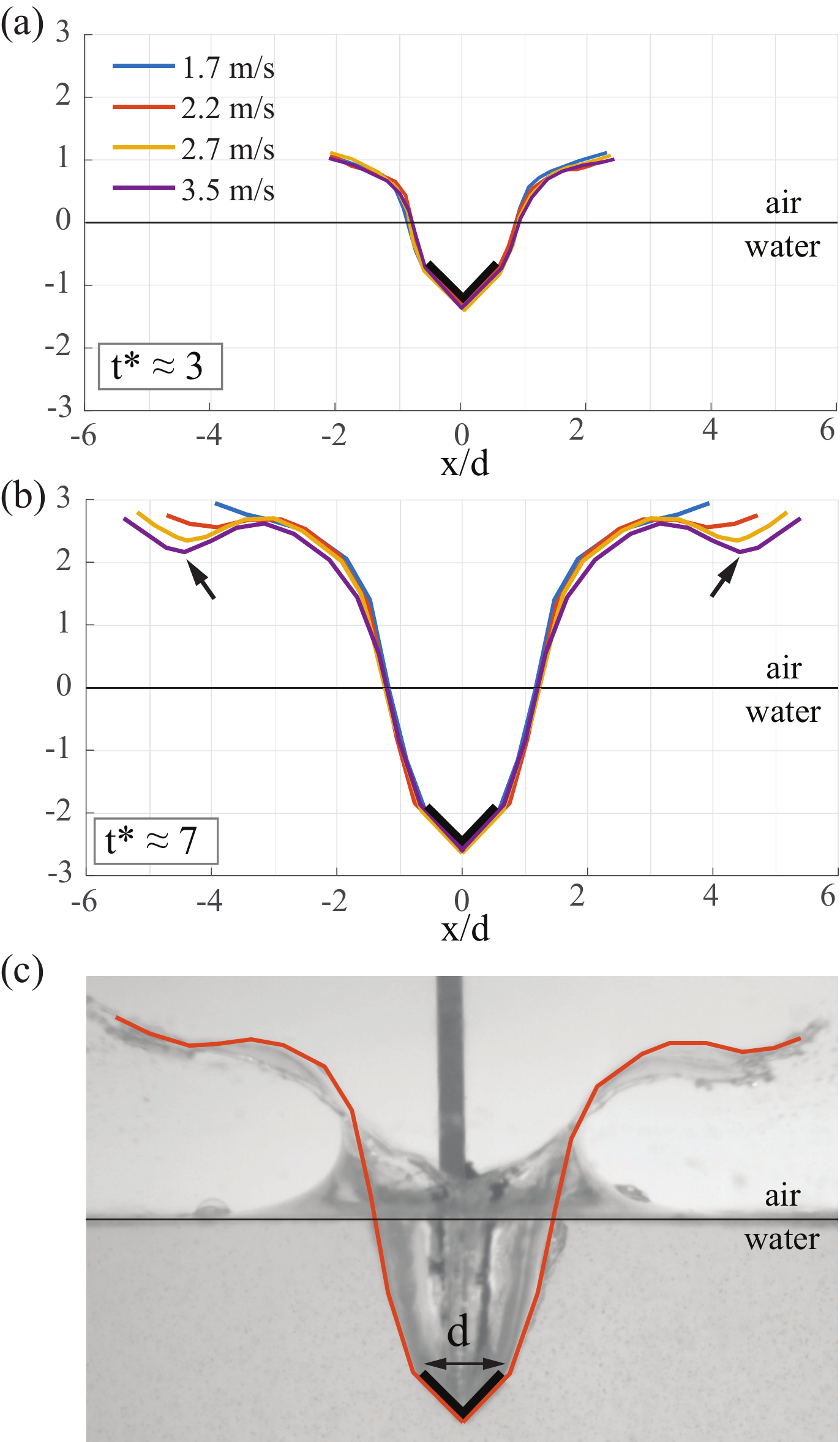} 
   \caption{(b) Shape reconstruction for $\alpha = 90 \degree$, at $t^* \approx 3$ (a) and  $t^* \approx 7$ (b), in the limit of vanishing gravity effects ($V = 1.7 - 2.5$ m/s, $\textrm{Fr} = 4 - 8$).  (c) Example of reconstruction for $V$ = 2.2 m/s ($\textrm{Fr} = 5.2$). The cavity is self-similar at all times, but the splash develop noticeable differences: as the impact velocity increases, the splash extend further and develop a double curvature. Notice the swelling of the dip (arrow) }
    \label{fig:splash_is_not_self-similar}
  \end{figure}


Snapshots of the splash and cavity shapes of a $90 \degree$ wedge entering water at various velocities $V$ are shown at two time instants  $t^* = 3$ and $t^* = 7$ in figure~\ref{fig:splash_is_not_self-similar}. At $t^* = 3$, the cavities and splashes corresponding to different velocities $V$ have the same form, but the splashes show notable differences at $t^* = 7$;  at larger entry velocities, splashes extend further and develop  a characteristic doubly-curved shape that we call ``arabesque". The fact that the splashes coincide well at short time $t^* = 3$ but not at longer time $t^* = 7$ suggests that these variations are not due to initial conditions. In order to investigate the physics underlying the splash evolution, we develop a one-dimensional model of the splash using first-principles and empirical observations.  
The model is based on the idea that the splash is primarily ballistic, and can be represented by a succession of discrete particles ejected at regular interval from the free water surface and moving under the influence of gravity, surface tension, and aerodynamic forces. Before we present the details of the models in~\S~\ref{sec:model_splash} and~\ref{sec:model_forces},
we examine the conditions of splash ejection  empirically.

\subsection{Splash ejection: empirical observations} \label{sec:splash_ejection}

\begin{figure}
\centering
\includegraphics[width=0.92\textwidth]{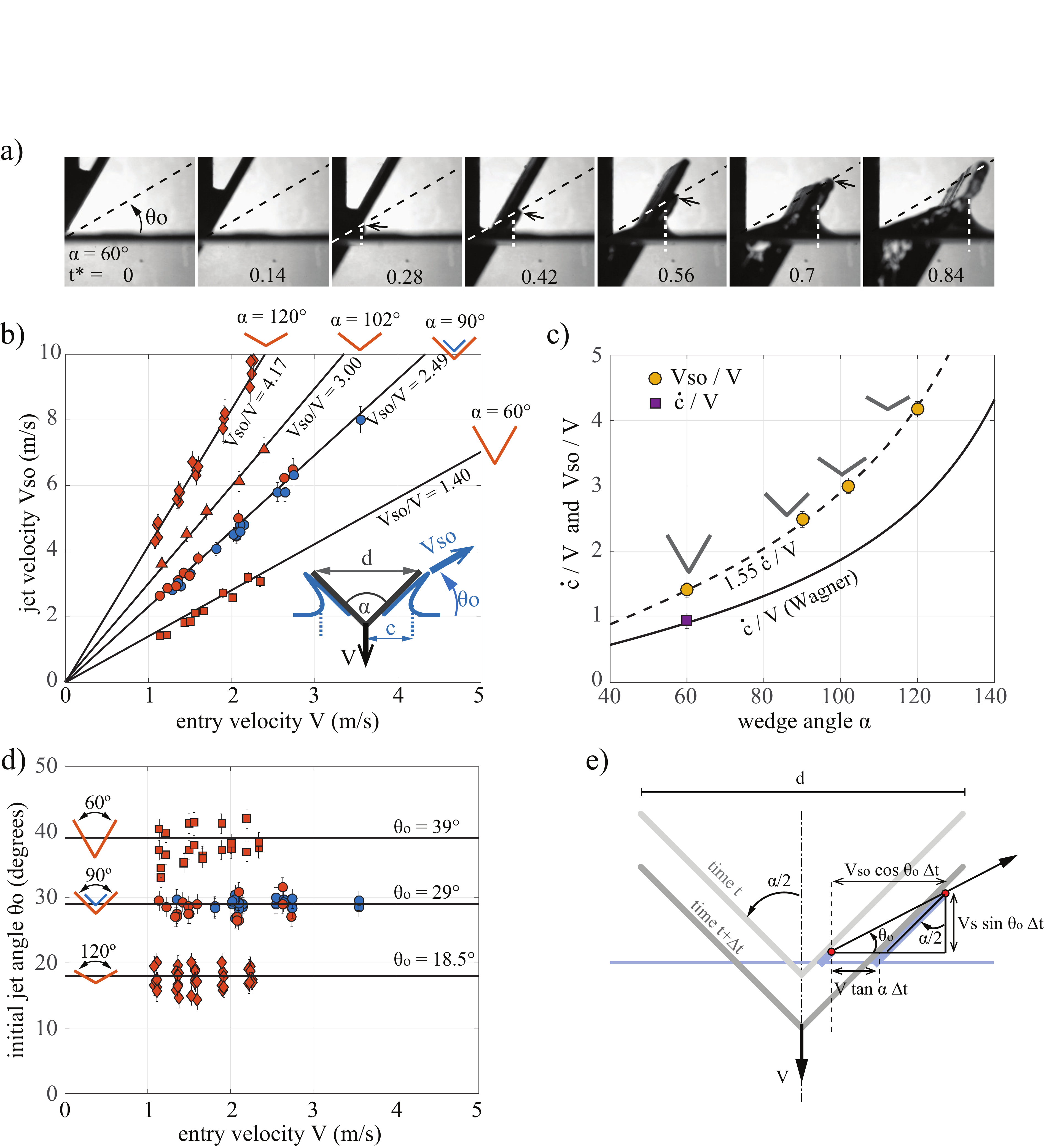}
\caption{Jet root expansion and jet velocity. Splash tip velocity $V_{\textrm{so}}$ is proportional to entry velocity $V$. The coefficient of proportionality increases with $\alpha$, and is measured to be $1.55 \times \dot{c}/V$. Vertical white dash line shows the location of the jet root; the black arrow shows the jet tip. Before submersion, shooting velocity and pitch angle are constant.}
\label{Fig:Vs_vs_V}
\end{figure}


Using high-speed video recordings, we systemically measure three relevant parameters of the splash kinematics at short times for various wedge angle $\alpha$: wetted length $c$, jet speed $V_{\textrm{so}}$, and jet angle $\theta_o$; see figure~\ref{Fig:Vs_vs_V}(a).

Raw measurements of $V_{\textrm{so}}$ as a function of entry velocity $V$ are shown in figure~\ref{Fig:Vs_vs_V}(b). Clearly, $V_{\textrm{so}}$ is linearly proportional to $V$, even at low entry speeds, and the coefficient of proportionality $V_{\textrm{so}}/V$ is an increasing function of $\alpha$. In figure~\ref{Fig:Vs_vs_V}(c), we depict $V_{\textrm{so}}/V$  versus $\alpha$, and compare it to the  expansion speed of the wetted length $\dot{c}/V$ obtained experimentally (purple square) and theoretically (solid line) based on equation (\ref{eq:c}). 
The speed of the splash tip $V_{\textrm{so}}/V$ is simply proportional to $\dot{c}/V$, with a coefficient of $1.55 \pm 0.05$ (dashed line), that is, 
\begin{equation}
\dfrac{V_{\textrm{so}}}{ V}  =  1.55 \left( \dfrac{\dot{c}}{V} \right) =  \dfrac{1.55 \pi}{2 \tan{\beta}}\label{eq:Vso} .
\end{equation}  
The jet angle $\theta_o$ between the trajectory of the water particles in the jet and the horizontal axis are reported in figure~\ref{Fig:Vs_vs_V}(d). 
Theoretical predictions of $\theta_o$ based on $V_{\textrm{so}}/V$ are obtained by assuming that the jet stays in contact with the wedge's surface, leading to the implicit equation for $\theta_o$,
\begin{equation}
\frac{\tan{\alpha}}{ \cos{\theta_o} - \sin{\theta_o} \tan{\alpha}} = \dfrac{V_{\textrm{so}}}{ V}   \label{eq:VsVs_vs_theta} .
\end{equation}  
Figure~\ref{Fig:Vs_vs_V}(d) shows that the theoretical predictions are in good agreement with the experimental measurements.
Taken together, these observations imply that the Wagner's approach allows us to predict  both the initial jet tip velocity $V_{\textrm{so}}$ and jet ejection angle $\theta_o$, given a single corrective parameter (the 1.55 constant) that is independent of $\alpha$. 
We are now equipped to tackle events past the initial submersion of the wedge, namely the long-term evolution of the splash shape. 


\subsection{Splash model: kinematics and initial conditions} \label{sec:model_splash}


\begin{figure}
   \centering
   \includegraphics[width=0.9\textwidth]{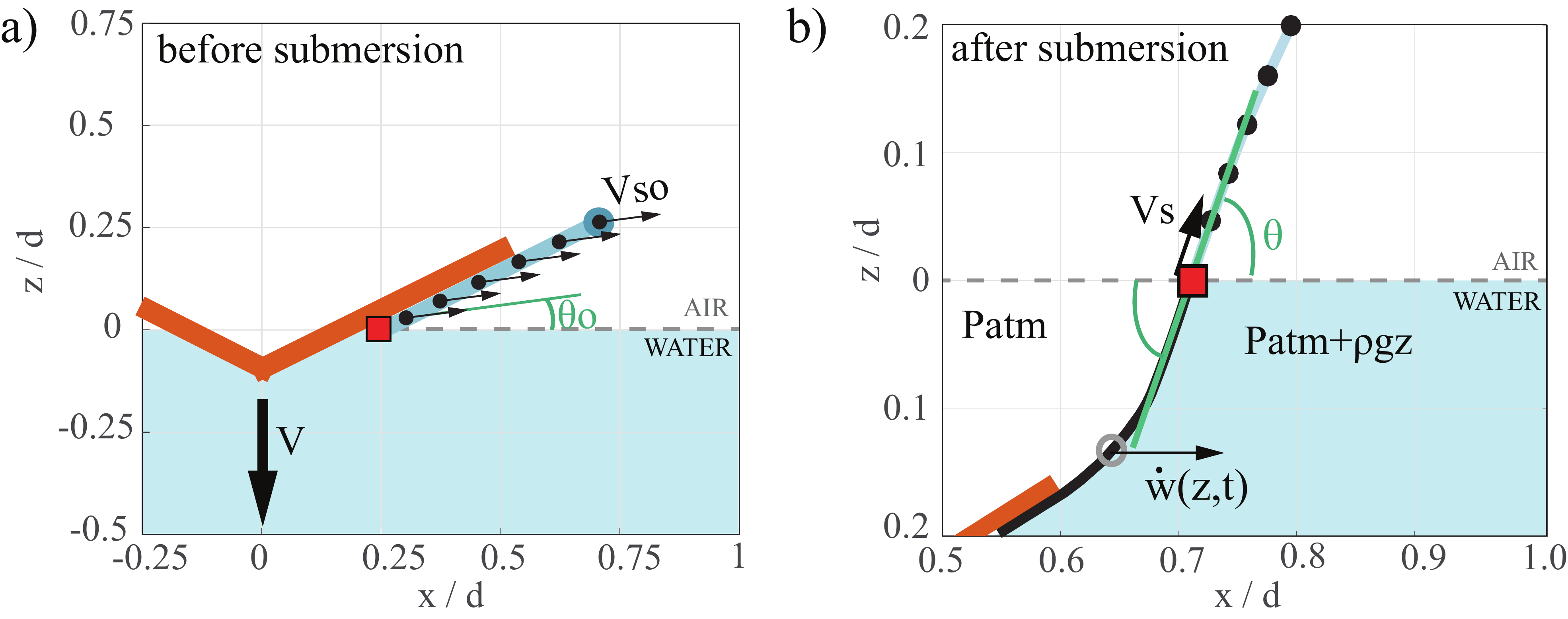} 
\caption{Initial conditions applied to shooting particles. (a) For times before wedge submersion, particles are shot with constant velocity $V_{so}$ and angle $\theta=\theta_0$. (b) After submersion, $\theta$ is adjusted to maintain continuity with the cavity shape, and the shooting velocity $V_s$ decreases exponentially.}
    \label{fig:model_initialconditions}
  \end{figure}

\begin{figure}
   \centering
   \includegraphics[width=0.9\textwidth]{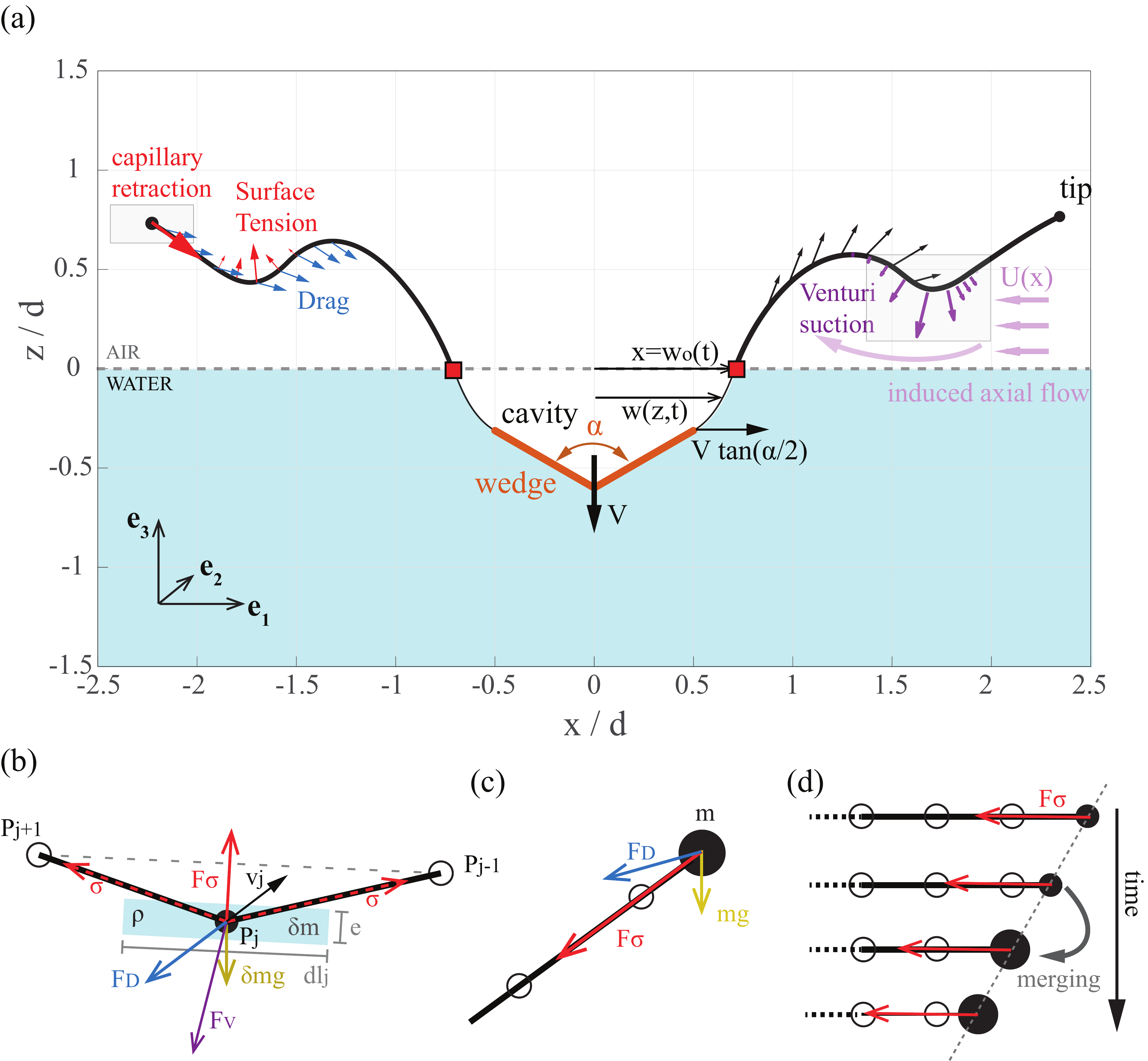} 
   \caption{Overview of the model and physical processes. The pitch angle is tied to the cavity local slope, and the shooting velocity decreases exponentially in time. Particles are subject to weight, drag, surface tension, and Venturi-induced suction. (b-d) Schematic representation of forces on an individual particle: (b) anywhere along the splash, and (c) at the tip, where capillary retraction takes place. (d) Initial stages of capillary retraction, showing the first particle merging. A straightforward merging routine allows the retraction speed to quickly converges to the Taylor-Culick velocity $V_\sigma = \sqrt {2 \sigma / \rho e}$, independently of the merging threshold.}
    \label{fig:model}
  \end{figure}



%
%

We develop a discrete fluid particles model of the splash, based on the idea that the splash is primarily ballistic. The splash is represented by a succession of discrete particles $P_j$, $j=1,2,3,\ldots \in \mathbb{N}$. Particle $j$ is located at $\mathbf{r}_j = x(P_j,t) \mathbf{e}_1 + z(P_j,t) \mathbf{e}_3$, where $P_j$ is a Lagrangian label of particle $j$, and it is moving at a velocity $\mathbf{v}_j = v_x(P_j,t) \mathbf{e}_1 + v_z(P_j,t) \mathbf{e}_3$ under the influence of gravity, surface tension, and aerodynamic forces.

Particles are initially ejected from the water surface ($z=0$)  at a shooting point $x(P_j,t_j)$, where $t_j$ represents the time of ejection of particle $j$. 
Our goal in this section is to determine the initial conditions, shooting position and velocity, for each particle. To this end, it is convenient to represent the velocity $\mathbf{v}_j$  of  particle $j$ by its speed $V_s(P_j,t)$ and pitch angle $\theta(P_j,t)$ measured from the horizontal.




Before the wedge is fully submerged, we make the assumption that the shooting point is located at  the intersection of the undisturbed water surface with the wedge given by $x  = V t_j \tan(\alpha)$, and all particles are assumed to eject with the same velocity $ V_{\textrm{so}}$ and angle $ \theta_o$ (figure~\ref{fig:model_initialconditions}(a)). That is to say, for ejection time $t_j \leq h/V$, the initial  conditions of particle $P_j$  are given by
\begin{equation}
V_s(P_j,t_j) = V_{\textrm{so}}, \quad \theta(P_j,t_j) = \theta_o, \quad x(P _j,t_j)=  V t_j \tan(\alpha).
\end{equation}
{This assumption is in agreement with the actual interface shape below the wedge (see figure~\ref{Fig:Vs_vs_V}) in that we consider the fluid pile-up as part of the splash.}

After submersion, $t_j \geq h/V$, the shooting point is located at the cavity wall $w(z=0)= w_o$, where $w$ is the cavity half-width and $w_o$ is the value of $w$ at the free water surface ($z=0$), see figure~\ref{fig:model}. 
 The initial shooting conditions are now dependent on the time of ejection $t_j$. Namely, the shooting velocity decreases with time and eventually decays to 0 while the base of the splash becomes steeper with time.
We consider the ejection velocity to decrease  exponentially such that
\begin{equation}
\label{eq:init_Vs_after_sub}
 V_s(P_j,t_j) = V_{\textrm{so}} e^{-(t_j - h/V) / \tau_s},
\end{equation}
where 
$\tau_s$ is a constant parameter that we set to $\tau_s = 0.025 / V$ regardless of the wedge angle. 

An implicit expression for the initial ejection angle $\theta(P_j,t_j)$ can be obtained by considering continuity between the slope of the splash and the slope of the cavity $w'_o = \left. d w /dz\right|_{z=0}$,
\begin{equation}
\label{eq:init_theta_after_sub}
              \dfrac{V_s(P_j,t_j) \sin({\theta}(P_j,t_j))}{V_s(P_j,t_j) \cos({\theta}(P_j,t_j)) - \dot{w_o}(t_j) } =  \tan(w'_o(t_j)).\\    
\end{equation}
Here,  $\dot{w_o}(t_j)$ and  $w'_o(t_j)$ are, respectively,  the horizontal speed and the slope of the cavity  at $z=0$ and $t= t_j$.

To close the system in~\eqref{eq:init_theta_after_sub} and~\eqref{eq:init_Vs_after_sub}, we need a model for the time evolution of the cavity wall $w$.    Here, we refer to~\cite{DuclauxClanet2007}, who applied a slice-averaged model to the transient dynamics of axisymmetric cavities created by spherical or cylindrical bodies. Using an unsteady potential flow model per slice, \cite{DuclauxClanet2007} found that the evolution of the cavity radius $R$ can be  described at each value of $z$ by the Rayleigh-Plesset equation: $R \ddot{R} + \frac{3}{2} \dot{R}^2 = \Delta p$, where $\Delta p = -gz$ is the pressure difference between the inside and the outside of the cavity at depth $z$ assuming atmospheric pressure inside the cavity. This equation is not directly applicable to wedges -- one would have to derive a new analytic expression for the slice-averaged flow potential assuming the wedge is infinitely long.
However, because our wedges have a relatively short aspect ratio $L/d$, we make the assumption that the equation derived in \cite{DuclauxClanet2007} applies in a modified form. Namely, we postulate that
\begin{equation}
w \ddot{w} + \frac{3}{2} C_c \dot{w}^2 = - gz \label{eq:cavity_EOM}
\end{equation}
where $C_c$ is an ad-hoc parameter (equal to 1 for axisymmetric cavities). By definition, $C_c$ is independent of $V$, but we expect it to depend on the geometry of the wedge, represented by the aspect ratio $L/d$ and opening  angle $\alpha$. The value of $C_c$ is adjusted for each wedge, by matching the cavity dynamics to high-speed experimental images. To obtain the cavity dynamics, we solve~(\ref{eq:cavity_EOM}) per slice for all slices between the top of the submerged wedge and the undisturbed water surface subject to initial conditions $w(t=0) = d/2$ and $\dot{w}(t=0) = V \tan \alpha$.

Put together,~\eqref{eq:init_Vs_after_sub}, \eqref{eq:init_theta_after_sub} and~\eqref{eq:cavity_EOM} form a closed set of equations that determine the initial conditions $V_s(P_j,t_j)$, $\theta(P_j,t_j)$ and $X_j$ for the ejection of particle $P_j$ for $t_j \geq h/V$.



\subsection{Splash model: force balance}
\label{sec:model_forces}

Particles are shot at regular intervals of time in such a way to be initially separated by a constant distance $d\ell_o$. The mass per unit length attributed to each particle is $\delta m = \rho e_o d\ell_o$,  where $e_o$ is the initial thickness of the splash sheet assumed to be constant during splash ejection. 
Each particle is subject to the following forces: weight $\delta m\,\mathbf{g}$,  drag $\mathbf{F_D}$, surface tension $\mathbf{F}_{\boldsymbol{\sigma}}$, and Venturi-induced suction $\mathbf{F_V}$. All forces are expressed per unit length of the splash sheet unless otherwise stated.
The force balance on particle $j$ is given by
\begin{equation}
\delta m \frac{d\mathbf{v}_j}{dt} = \delta m \, \mathbf{g}+  \mathbf{F_D} + \mathbf{F}_{\boldsymbol{\sigma}} + \mathbf{F_V}. \label{eq:OEM}
\end{equation}
{Our model bears similarity with previous models developed in the context of moving fluid sheet \cite{Gart2013}. But it goes further in that it takes into account the sheet stretching and includes original contributions such as Venturi-induced suction.}
Drag $\mathbf{F_D}$ on particle $j$  is calculated by considering the splash segment $\Delta \mathbf{r}_j = \mathbf{r}_{j-1} - \mathbf{r}_{j+1}$ of   length $d\ell_j = \| \Delta \mathbf{r}_j \|$, moving with velocity $\mathbf{v}_j$ at an angle of attack $\psi_j$ defined as the angle between $\mathbf{v}_j$ and $\Delta \mathbf{r}_j$. This leads to  \cite{AndersenWang2005}
\begin{equation}
\label{eq:drag}
\mathbf{F_D} = - \frac{1}{2} C_D \sin^2(\psi_j) \rho_a d\ell_j  \| \mathbf{v}_j \|  \mathbf{v}_j . 
\end{equation}
Here, $C_D$ 
designates the drag coefficient for when the element is perpendicular to the incoming flow, which we fix at $C_D = 2.5$ for all experiments. {The drag force is  calculated using the   absolute traveling velocity of the fluid particle, and not its  velocity relative to the ambient air. The effect of the air flow is considered independently in the Venturi-suction force made explicit later on in this section}.





Surface tension is accounted for in the most straightforward way: each particle experience a longitudinal traction of  magnitude $\sigma$ from its closest neighbors, as represented in figure~\ref{fig:model}(b). For $j\neq 1$, one has
\begin{equation}
\label{eq:surfacetension}
\mathbf{F}_{\boldsymbol{\sigma}} = \sigma \left( \frac{\mathbf{r}_{j-1} - \mathbf{r}_{j}}{\| \mathbf{r}_{j-1} - \mathbf{r}_{j} \|} + \frac{\mathbf{r}_{j} - \mathbf{r}_{j+1}}{\| \mathbf{r}_{j} - \mathbf{r}_{j+1} \|} \right).
\end{equation}
This force is normal to the local tangent $\Delta \mathbf{r}_j $, and acts as a restoring force: it tends to cancel any shape curvature. The particle at the tip ($j=1$) is a particular case: surface tension results in only one longitudinal force directed towards its next neighbors, leading to the retraction of particle at the tip (figure~\ref{fig:model}(c)). This retraction is observed in systems such as free sheets and ligaments \cite{Marmottant2004,Lhuissier2009}. The speed of retraction, relative to the sheet, is constant and known as the Taylor-Culick velocity $V_\sigma = \sqrt {2 \sigma / \rho h}$ \cite{Taylor1959,Culick1960}. To account for capillary retraction in our model, the receding particle of initial mass $\delta m$ has to merge with its successive closer neighbors, as shown in figure~\ref{fig:model}(c). When merging happens, (i) the new particle's mass increases by $\delta m$, (ii) the new particle is positioned at the center of mass of the two former particles, and (iii) the new particle's momentum is the sum of the momentum of the the two former ones. Given these conditions, we observe that the retraction speed of the tip particle quickly converged to $V_\sigma$, independently of the merging criteria (inter-particle distance). 


\begin{figure}
   \centering
   \includegraphics[width=0.99\textwidth]{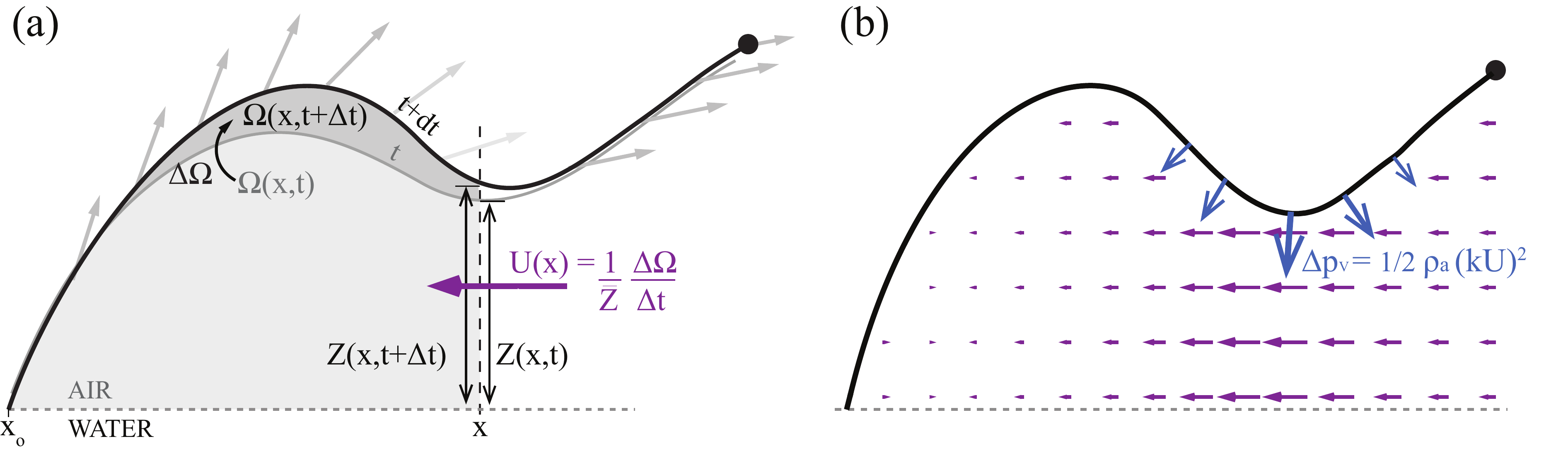} 
   \caption{(a) Average axial velocity into the chamber is determined using volume conservation. (b) Typical calculated axial flow field (purple), and resulting suction force on the sheet (blue).}
    \label{fig:Venturi_volumechange}
  \end{figure}
The last force considered in this model, that has a significant effect on the late-stage development of the sheet, is a Venturi-induced suction. This suction happens because the motion of the splash dome induces an axial flow of air rushing in to fill the expanding chamber beneath it, as represented in figure~\ref{fig:Venturi_volumechange}. The air is accelerated through the gap between  the undisturbed  water surface and the lower part of the splash, causing the pressure to drop and generating a downward suction. To calculate this suction force, we begin by evaluating the volume
$\Omega(x,t)$ of the chamber under the splash
\begin{equation}
\Omega (x,t) = \int_{x_{{o}}}^{x} Z(\tilde{x},t) d\tilde{x},
\end{equation} 
where $x_{{o}}$ is the position of the base of the splash and $Z(x,t)$ refers to the local height of the splash above the free water surface. Assuming the chamber is two-dimensional, that is, no flow in the direction perpendicular to the drawing plane,
 the average volume change
$\Delta \Omega(x,t) =  \Omega (x,t+\Delta t) - \Omega (x,t)$ of  the splash is related to the average flow $U(x,t)$ passing through the section $\bar{Z}= \frac{1}{2} [ Z(x,t) + Z(x,t+\Delta t) ]$, yielding
\begin{equation}
U = \dfrac{1}{ \bar{Z}}\dfrac{\Delta \Omega}{ \Delta t}. 
\end{equation}
Using Bernoulli's principle {between a position far away from the splash} and the bottom of the sheet, we can estimate the pressure drop as $\Delta p_{\textrm{v}} = - \frac{1}{2} \rho_a (k {U})^2$, where $k U$ is the corrected axial velocity. The correction factor $k$ accounts for the difference between the real system and the idealized model, including non-uniformity of this velocity in the vertical direction and three-dimensional effects. {We expect $k$ to depend on $\alpha$. In particular, $k$ should converge towards $1$ as $\alpha$ increases because the splash flies lower and the lubrication approximation becomes therefore more justified. We adopted $k=1.5$ for all computations presented in section~\ref{sec:120deg_splash} ($\alpha = 120 \degree$).}

The magnitude of the suction force $\mathbf{F_V}$ at particle $P_j$ is given by $\Delta p_{\textrm{v}} d\ell_j$, where $\Delta p_{\textrm{v}}$ is 
evaluated at $x = x(P_j,t)$ and $d\ell_j = \| \Delta \mathbf{r}_j \|$ as previously defined; The direction of  $\mathbf{F_V}$ is along the local normal $\Delta \mathbf{r}_j^\perp / \| \Delta \mathbf{r}_j \| $ to the sheet, namely,
\begin{equation}
\label{eq:suction}
\mathbf{F_V} = - \frac{1}{2} \rho_a (k {U})^2 \Delta \mathbf{r}_j^\perp.
\end{equation}

We substitute expressions~\eqref{eq:drag},~\eqref{eq:surfacetension} and~\eqref{eq:suction} for the forces due to drag, surface tension and Venturi-induced suction into~\eqref{eq:OEM}. The resulting equations are integrated using an explicit forward Euler method with typical time step $dt = 10^{-6}$. The initial distance between particles $d\ell_o= 100$ $\mu$m is chosen to be of the order of the sheet thickness.




\section{Splash Evolution} \label{sec:120deg_splash}

\begin{figure}
   \centering
   \includegraphics[width=0.99\textwidth]{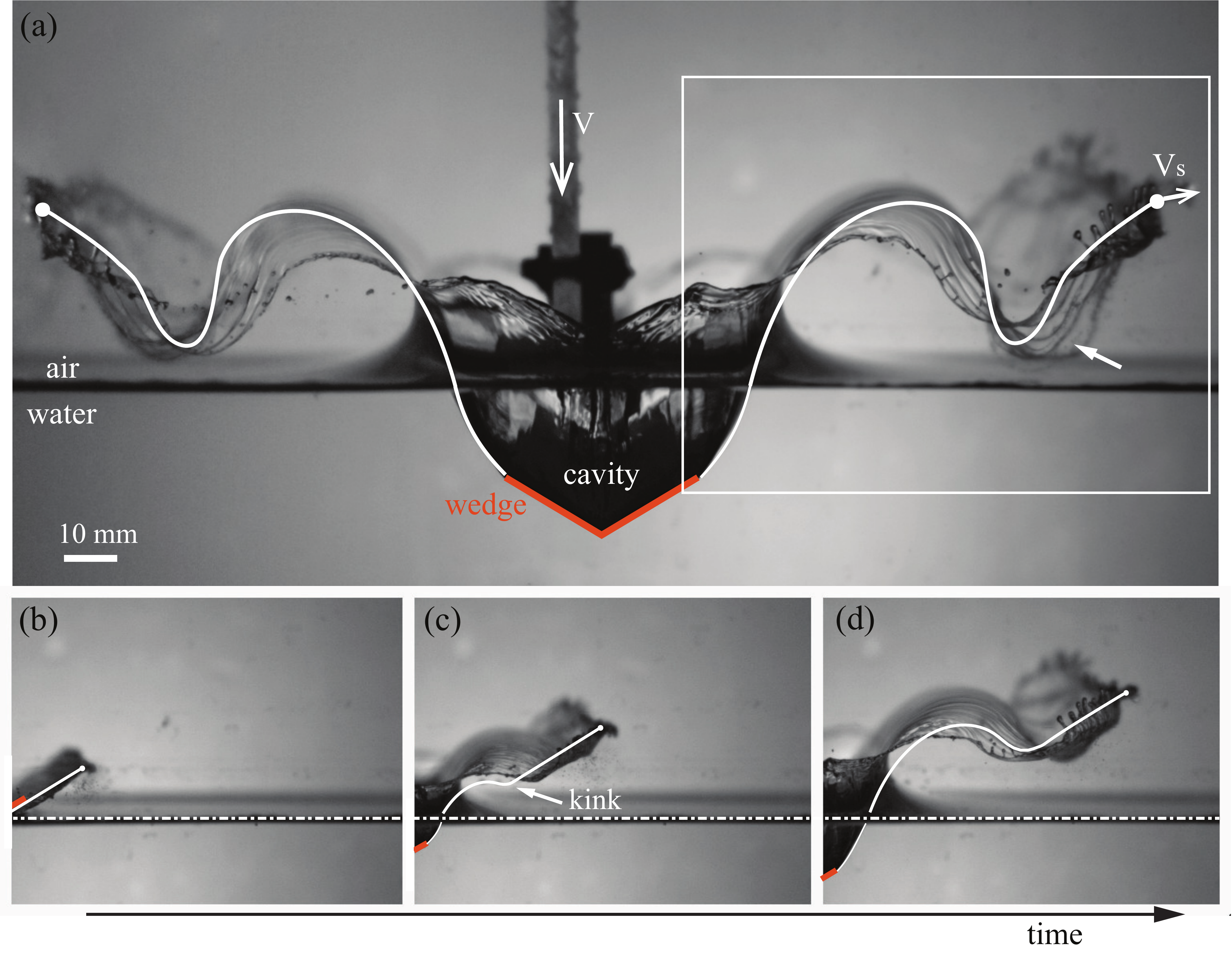} 
   \caption{(a) Snapshot taken 15 ms after impact of a $\alpha = 120 \degree$ wedge in water at $V = 1.92$ m/s, and (b-d) preceding time-evolution of the splash sheet; snapshots are $3.1$ ms apart. The leading part of the sheet is ejected with a speed $V_s \approx 4.5$ m/s. Note the strong downward suction of part of the splash sheet (white arrows) leading to its rapid fragmentation. Computed shapes are superimposed in white.}
    \label{fig:120splash_sequence}
  \end{figure}

Figure~\ref {fig:120splash_sequence} presents the evolution of the splash following the impact of a 120$\degree$ wedge in water at $V = 1.92$ m/s. The ejecta deforms into a fairly complex shape, characterized by two inversions of curvature from its base to its rim. Of particular interest is the formation of a dip, resulting from the strong downward pull of the thinnest and lowest part of the splash. As we demonstrate in the following, the existence of this singular feature is tied to two keys ingredients: the generation of a kink at short times and the growth of this kink, favored by aerodynamics and hindered by surface tension (figure~\ref {fig:120splash_sequence}(b-d)). We also show that for moderate and high wedge angles, Venturi effect is the dominant aerodynamic force driving the splash deformation. We illustrate the physical concepts with experiments corresponding to a particular case: a 120$\degree$ wedge of width $d = 36$ mm. However, we emphasize that the model and all discussions apply to wedges of various opening angle and aspect ratio.  

\begin{figure}
   \centering
   \includegraphics[width=0.99\textwidth]{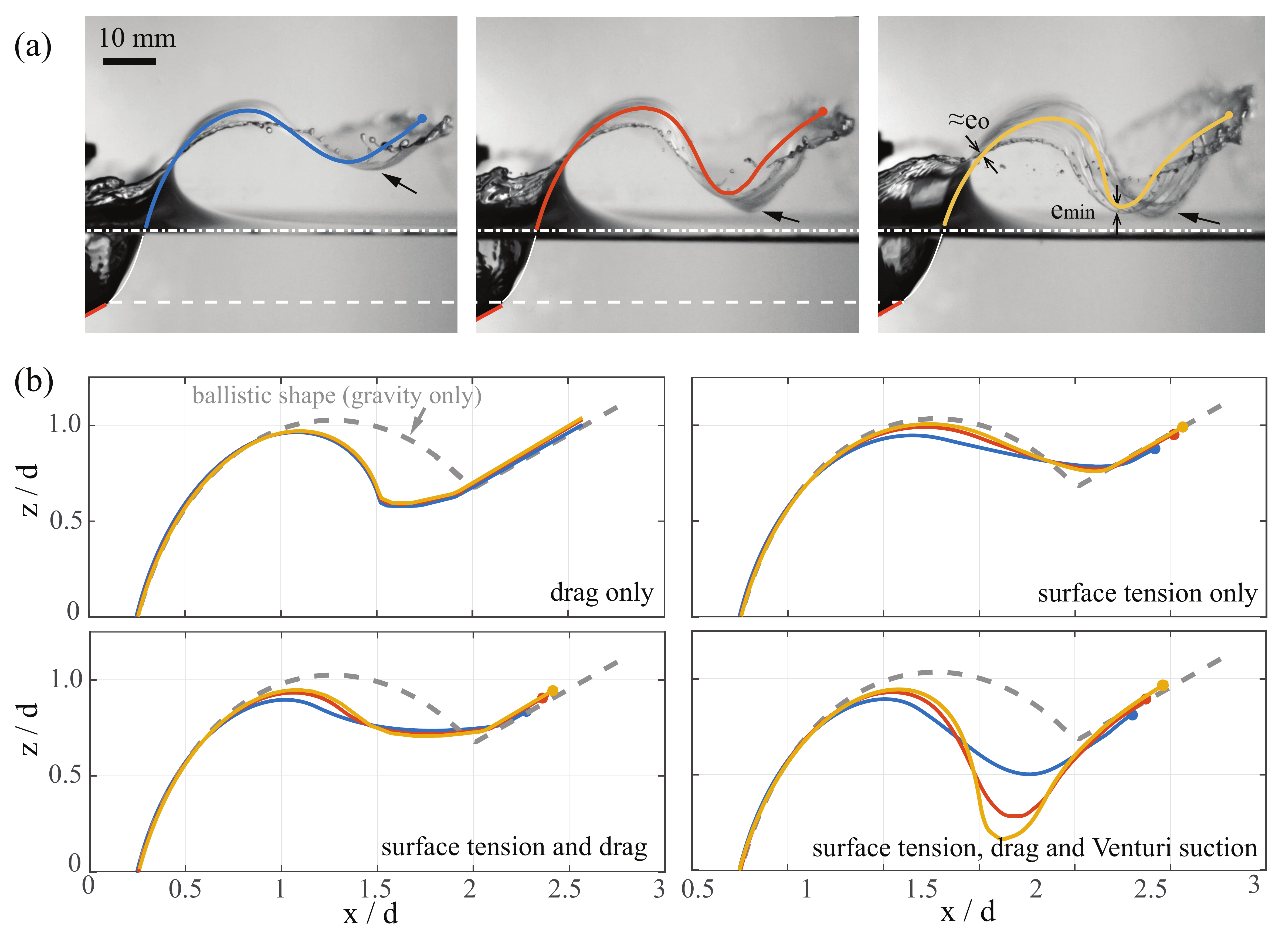} 
   \caption{(a) Experimental snapshots and corresponding computed shapes taken at $t^* = t / (h/V) \approx 2.8$ after impact for increasing impact velocity $V$: $1.11$~m/s - $1.53$~m/s - $1.92$~m/s. Note that the dip in the splash sheet (black arrow) gradually swells as velocity increases. The third panel shows the last computable shape at $t^* \approx 2.6$, for which the thinning ratio is $e_{min} / e_o \approx 1/30$. (b) Quantification of the effect of the various forces. While neither drag or surface tension can account for the bending, Venturi-mediated suction does. }
    \label{fig:snapshots}
  \end{figure}

To highlight the effect of entry velocity, we reproduce in figure~\ref {fig:snapshots}(a) three snapshots for gradually increasing entry velocity $V= 1.11$, $1.53$, and $1.92$ m/s, taken at the same dimensionless time $t^* = 2.6$. The snapshots are similar in many ways. Looking below the free water surface, the wedge's penetration and cavity shapes are virtually the same. Above the free surface, the splash shapes share some similarities. For instance, the sheet rim, at the far right of each picture, is nearly at the same location.  There is, however, one major difference between the three snapshots: the depth of the depression. In the left panel, the lowest part of the dip is at about half the height of the dome, while in the right panel, the depression nearly reaches the water surface. This difference can be, erroneously, attributed to the effect of drag on the fast-moving sheet. We shall show that this idea is essentially wrong, and while drag does have a significant effect on the water sheet, it only contributes little to the strong downward pull the dip of the splash sheet is subject to. Figure~\ref {fig:snapshots}(b) uses our model to quantify the effect of various forces on the splash shape. Drag and surface tension, alone or combined, generate splash shapes that are nearly identical at different entry velocities (labeled by color).  Venturi suction is needed to be able to reproduce correctly the observed shapes and variation with $V$: the  splash shapes, computed with all forces, are superimposed onto figure~\ref {fig:snapshots}(a), showing excellent agreement between the 1D model and the experimental observations. 

\begin{figure}
   \centering
   \includegraphics[width=0.99\textwidth]{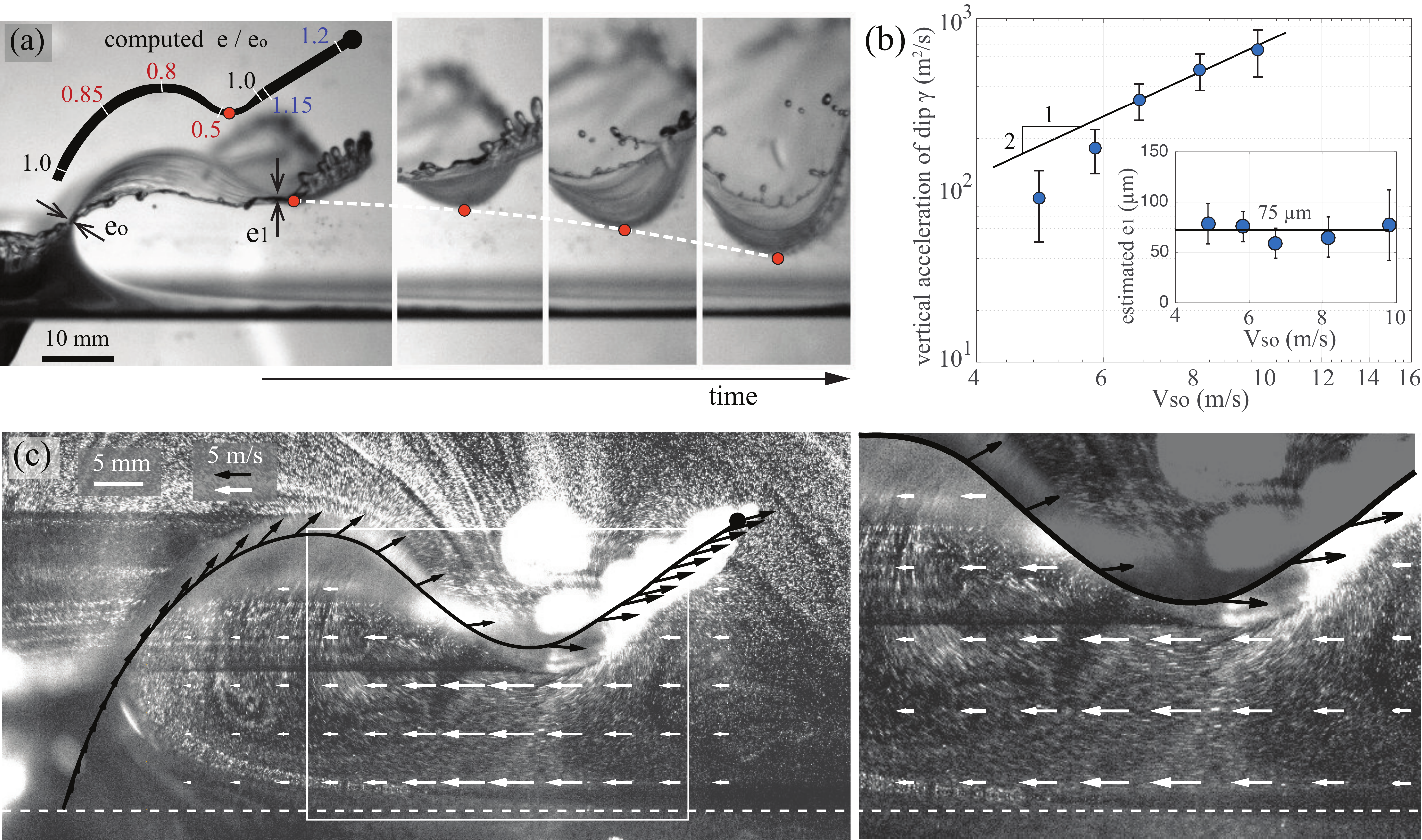} 
   \caption{(a-b) Quantification of the strong downward pull on the thinnest ($e / e_o \approx 0.5$) part of splash sheet. When surface tension effects are negligible, the acceleration is proportional to $V_{\textrm{so}}^2$, suggesting a purely aerodynamic force. Assuming the associated aerodynamic pressure scales as $\frac{1}{2} \rho_a V_s^2$ leads to an indirect estimate of the sheet thickness $e_o$. (c) Visualization of the fluid motion around the splash at $t^*=0.7$ for $V = 1.36$ m/s using oil droplets, highlighting the axial flow into the expanding chamber beneath the splash sheet. The estimated average axial velocity (white) from mass conservation show good agreement with the experiment. The splash sheet and the velocity of individual particles within the sheet as shown in black.}
    \label{fig:dip_acc}
  \end{figure}

\subsection{Splash depression}

In order to elucidate the nature of the vertical force causing the depression in the splash sheet, we examine the dynamics of the dip. In figure~\ref {fig:dip_acc}, we measure the vertical acceleration $\gamma$ of a marker located at the local minima in the dip, designated by a red dot. The measurements are taken shortly after the downward dip becomes clear, in the time interval between the first two snapshots of figure~\ref {fig:dip_acc}. We find that acceleration is an increasing function of the splash velocity $V_s$ and, consequently, of $V$, with values well above the gravitational acceleration: around $10$ $g$ for the slowest velocity, up to $90$ $g$ for $V_s = 2.35$ m/s. More interestingly, we find that for $V_s > 7$, $\gamma$ is proportional to $V_s^2$, suggesting the downward force has an aerodynamic origin. 

We push the analysis further in order to determine the local thickness $e_1$ (and, thus, $e_o$) and to derive a criterion for the growth of the dip. To this end, we write the force balance in the vertical direction on a small fluid element of length $d\ell$, width $L$, and thickness $e_1$. The forces are drag $F_D$, surface tension $F_\sigma$ acting upwards, and an aerodynamically-related downward suction $F_V = \frac{1}{2} \rho_a d\ell L V_s^2$. We ignore the drag contribution because the vertical velocity of the marker is at least 5 times smaller than $V_s$. We are left with the force balance in the vertical direction 
$\rho e_1 L  d\ell  \gamma=  F_\sigma - F_V$.
Dividing throughout by $L d\ell$, we get the balance law in terms of pressure difference between  a capillary contribution $ \Delta p_\sigma$ and an aerodynamic contribution $\Delta p_V = \frac{1}{2} \rho_a V_s^2$, 
\begin{equation}
 \rho e_1 \gamma = - \Delta p_\sigma +  \Delta p_V.
  \label{eq:balance_dl_pressure}
\end{equation}
From~\eqref{eq:balance_dl_pressure}, we expect the growth of the dip to be inhibited when the restoring effect of surface tension overcome the destabilizing effect of the suction pressure. 

To derive a criterion for predicting the dip growth, we assume that the local radius of curvature of the splash scales as the wedge's lateral size $d$, in qualitative agreement with observation. The capillary pressure contribution is thus $\Delta p_\sigma \approx 2 \sigma / d$ across the two interfaces. Substituting into~\eqref{eq:balance_dl_pressure}, we get the following criteria for the growth of the dip: $\frac{1}{2} \rho_a V_{\textrm{so}}^2 > 2 \sigma / d$. For $d = 36$~mm and $\rho_a = 1.25$ Kg/m$^3$, the criteria yields $V_{\textrm{so}} > 2.5$ m/s. Our experimental trials correspond to $V_{\textrm{so}}$ between 4 and 10 m/s, for which we always see dip growth, as expected. This criterion corresponds to $\textrm{We}_a > 4$ when rewritten in terms of a modified Weber number $\textrm{We}_a =  \rho_a V_{\textrm{so}}^2 d / \sigma$.   Unlike the Weber number commonly used in impact problems, $\textrm{We}_a$ depends on the density of air instead of water. 

Equation~\eqref{eq:balance_dl_pressure} also yields an estimate of the local thickness $e_1 = ( \frac{1}{2} \rho_a V_{\textrm{so}}^2 - 2 \sigma / d) / \rho \gamma$. A value of $e_1\approx 75$ $\mu$m is obtained based on this expression; see inset of figure~\ref {fig:dip_acc}(b).  According to figure~\ref{fig:dip_acc}(a), we consider $e_1 / e_o \approx 0.5$ and get that $e_o \approx 150$ $\mu$m, a value that we adopted for all computations presented in this work.

\subsection{Effect of non-dimensional parameters on splash shape}

\begin{figure}
   \centering
   \includegraphics[width=0.7\textwidth]{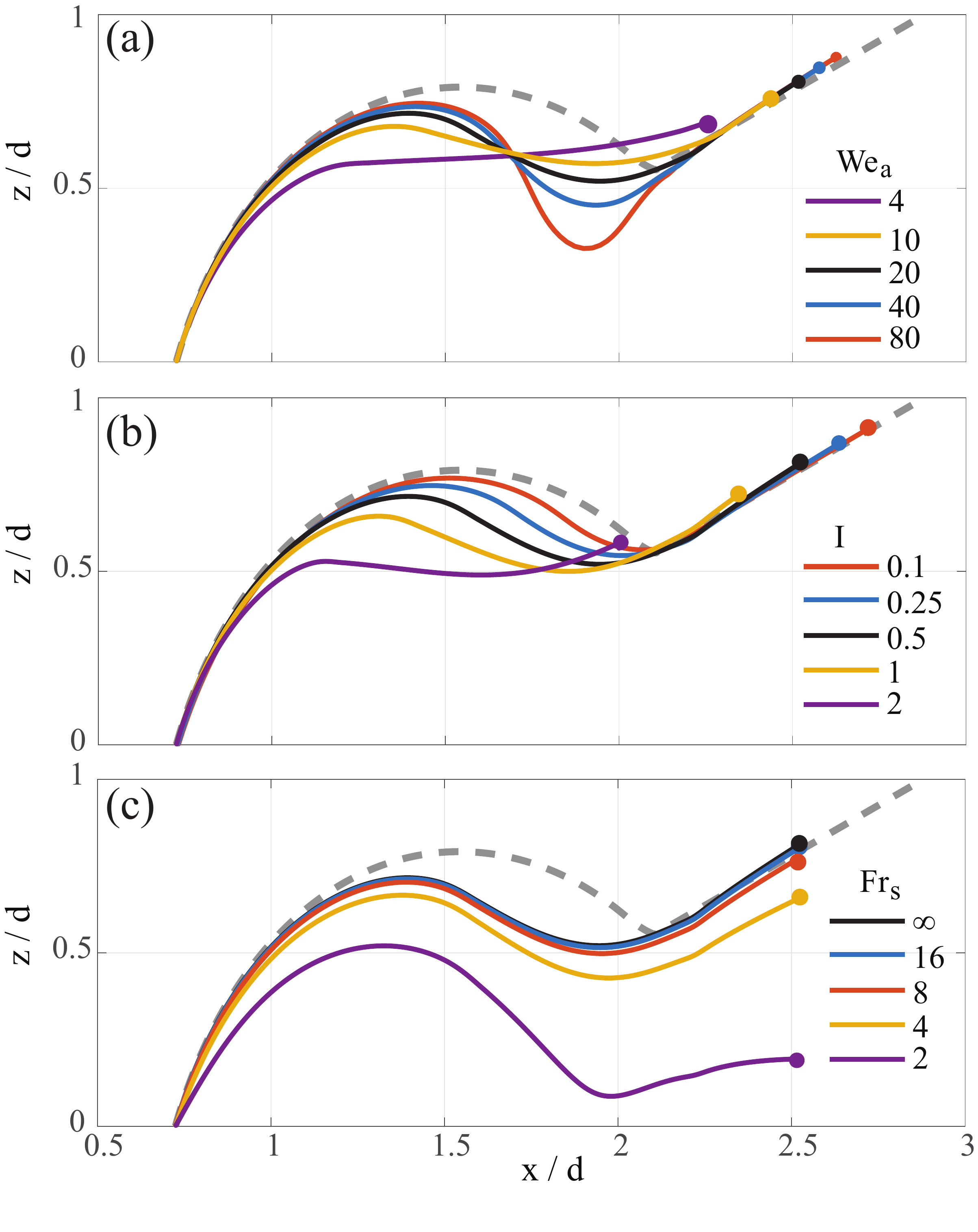} 
   \caption{Snapshots taken at $t  = 2.5$ showing the effect of the three non-dimensional parameters affecting the splash shape: We$_a = \rho_a V_{\textrm{so}}^2 d  / \sigma$, $I = C_D (\rho_a / \rho) d / e_o $ and Fr$_{s}= V_{\textrm{so}} / \sqrt{g d}$.}
    \label{fig:effect_of_Fr-I-Fr}
  \end{figure}

The modified Weber number $\textrm{We}_a$ reflects the competition between surface tension and aerodynamic suction. In figure \ref{fig:effect_of_Fr-I-Fr}(a) is a depiction of the expected splash shapes for various $\textrm{We}_a$. For low $\textrm{We}_a$, the splash is shorter due to capillary retraction, and flatter because surface tension is preventing bending. For higher $\textrm{We}_a$ numbers, the bending become more pronounced and the splash shapes reach further.  


There are, however, two other effects that the splash shape depends on: drag and gravity. The primary effect of drag is to slow down the splash, in a fairly uniform fashion. One way to estimate the effect of drag on the splash shape is to calculate the relative deceleration of a particle along its trajectory. Let's consider the motion of a slice of splash, of mass $\delta m = \rho e_o d\ell$, traveling at velocity $v$, over a distance proportional $d$, and subject to a drag force $F_D = \frac{1}{2} C_D \rho_a d\ell v^2$. According to Newton's second law of motion, the deceleration can be approximated by
$\rho e_o \Delta v /\Delta t =  \frac{1}{2} C_D {\rho_a} {v^2}$, which leads to, upon substituting $\Delta t \approx d / v$, 
\begin{equation}
\frac{\Delta v}{v} \approx \frac{1}{2} C_D\frac{\rho_a}{\rho} \frac{d}{e_o} 
\end{equation}
The quantity $I$ is independent of $v$ and thus of the entry velocity $V$, emphasizing that drag is not responsible for the various degrees of splash bending observed for different entry velocities $V$. However, the shape of the splash is affected by changing $I$, either by varying the air-water densities or by changing the dimensions of the wedge. Figure \ref{fig:effect_of_Fr-I-Fr}(b) shows the effect of $I$ on the splash shape. Smaller $I$ (thinner or lighter sheet) produces shorter splashes, without affecting the shape of the splash  itself.

Lastly, we redefine a Froude number based on the splash velocity, $\textrm{Fr}_{s}= V_{\textrm{so}} / \sqrt{g d}$, which reflects the competition between aerodynamic forces  and gravity.  Figure \ref{fig:effect_of_Fr-I-Fr}(c) shows that decreasing $\textrm{Fr}_{s}$ (increasing gravity) induces a global downward  motion and tilting of the whole splash. Significant differences are expected starting at $\textrm{Fr}_s = 4$. The experiments presented before are in the range $\textrm{Fr}_s = 7.7 -  15$, confirming gravity has indeed negligible impact on the splash shape.

\subsection{Sheet fragmentation}

\begin{figure}
   \centering
   \includegraphics[width=0.9\textwidth]{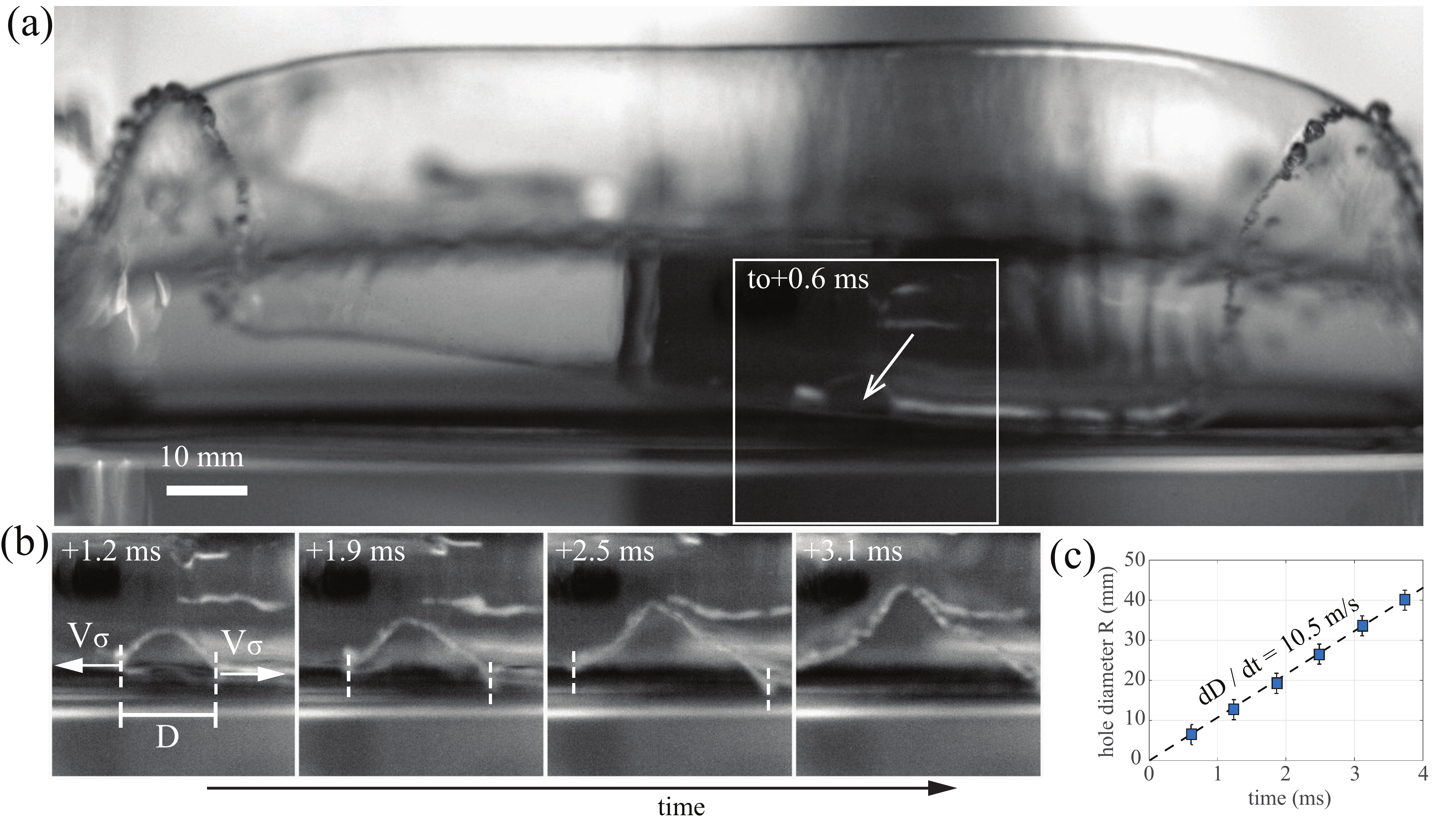} 
   \caption{(a-b) Front view of the splash at $t^* \approx 0.85$ for $V = 1.15$ m/s showing the spontaneous opening of a hole on the thinnest part of the dip. (c) The hole expansion velocity is $2 V_\sigma$, where $V_\sigma = \sqrt {2 \sigma / \rho_a e_{min}}$ is the Taylor-Culick velocity, and provides an indirect measurement of the local sheet thickness $e_{min} \approx 5$ $\mu$m, $30$ times smaller than the estimated initial sheet thickness $e_o$.}
    \label{fig:hole_in_dip}
  \end{figure}

The strong downward motion of the dip has stretches the water sheet. To estimate the stretching ratio, given the initial sheet thickness that we estimated in the previous section ($e_o = 150$ $\mu$m), we need an estimate of the sheet thickness $e_{min}$ of the splash sheet at its lowest and thinnest point. Fortunately, we have access to $e_{min}$ by observing a spontaneous puncture of the sheet. Figure~\ref {fig:hole_in_dip} presents the time sequence of the expansion of a hole in the bottom of the sheet at $t^* \approx 2.5$. The hole expands in all directions at a velocity $V_\sigma = \sqrt {2 \sigma / \rho e_{min}}$, known as the Taylor-Culick velocity \cite{Taylor1959,Culick1960}. 
We measure $V_\sigma = 5.25 \pm 0.05$ m/s
and calculate $e_{min} = 5.1 \pm 0.3$ $\mu$m. This is about 30 times smaller than the estimated $e_o$, in agreement with the maximum stretching ratio of last computable shapes of figure \ref{fig:snapshots}. More importantly, it underlines the dramatic stretching induced by the the downward suction. These calculations are the first steps towards developing a low-order model of fragmentation to be pursued in future work.


\section{Conclusion}
We considered the dynamics of diving wedges. In the first part, we studied the force applied to the wedges during entry. We showed that while sharp wedges enter the water smoothly, obtuse wedges experience a large transient peak force before total submersion. The transition between smooth and impactful entry happens for $\alpha \simeq 70 \degree$. 
Our experimental measurements of the maximum force and the time of occurrence compare well with existing impact force theories \cite{Logvinovich1972}, after incorporating corrections due to finite aspect ratio $L/d$ of the wedge~\cite{Meyerhoff1970}. We also showed that, after submersion, diving wedges are subject to smaller drag forces, about two-fold smaller, than the drag forces on immersed wedges. We show that this difference is due to the presence of the cavity, and that the magnitude of the drag is well predicted using existing cavitation theory, even though the origin of the cavity is different.

The second part of this study focused on the dynamics of the splash. We showed that while the velocity at entry doesn't have any appreciable effect on the shape of the cavity, as noted by previous authors \cite{DuclauxClanet2007}, it does have a significant effect on the splash shape. Large velocities generally lead to increasingly more ample and further reaching arabesques. We proposed a 1D model of the splash, taking into account the physical forces acting on the ejected water sheet. We identified a Venturi suction force, because of the air rushing in between the sheet and the water surface. This phenomenon is similar to the one observed by \cite{Thoroddsen2011} for impact of drops in water, but at a much larger scale and smaller velocity, allowing better visualization and seeding. The shape of the splash is driven by a competition between Venturi suction, driving the instability, and surface tension that acts as a restoring force. For low-flying splashes, namely those created by large wedge angles, the Venturi suction overcomes surface tension and the splash sheet collapse onto the water surface.


The 1D splash model satisfactorily captures the splash development, but it can be improved in several ways. First and foremost, one can take into account the water pile-up under the wedge. The main change would be that the occurrence of full submergence will happen quicker, and as a result, the outermost, straight portion of the splash will be shorter, most likely improving the model's fidelity. To complete this approach, one would have to consider the pressure profile on the wedge to infer the initial conditions given to the water particle until the wedge is fully submerged with respect to the undisturbed surface. Although harder to implement, this method would bridge the gap between the pressure profile and splash shape: the kink in the splash shape would be expected to form shortly after the jet root escapes the wedge's edge, because of the large pressure in this portion of the wetted region. 

Another direction to improve the model is to account for the thickness of the splash and its dynamic evolution.
Some impact models suggest that thickness (measured at the jet root) is solely a function of the size and opening angle of the wedge \cite{Pierson1950}, while the thickness scale $\sqrt{\nu D / V_s}$, commonly used in sphere impact problems, include both splash velocity and viscosity \cite{Thoroddsen2011}. Additionally, the thickness is expected to vary slowly along the splash sheet, being thinner at the tip area (excluding the rim), and thicker closer to the base. To our knowledge, little is known about this dependence, especially after the jet root has escaped. A better understanding of the mechanisms governing the splash thickness would form a first step towards developing a fragmentation theory.


\paragraph{Acknowledgment.}  The work of   L. Vincent, T. Xiao, D. Yohann and E. Kanso is partially supported by the NSF CBET grant 1512192 (to E. Kanso).


\bibliographystyle{abbrv}
\bibliography{biblio_crossingtheboundary}

\end{document}